\documentclass[12pt,english]{article}
\pdfoutput=1
\usepackage[T1]{fontenc}
\usepackage{amssymb}
\usepackage{amsmath}
\usepackage{cite}
\usepackage{epsfig}
\usepackage{scalerel}
\usepackage[unicode=true,bookmarks=true,bookmarksnumbered=false,bookmarksopen=false,
 breaklinks=false,pdfborder={0 0 1},backref=false,colorlinks=true]
 {hyperref}
\usepackage[hang,flushmargin]{footmisc}
\usepackage{color}
\usepackage{shuffle}
\usepackage{cleveref}
\usepackage{nicematrix}
\usepackage{todonotes}
\usepackage{booktabs} 		
\usepackage{geometry}

\usepackage[latin1]{inputenc}
\usepackage{amsmath}
\usepackage{amsfonts}
\usepackage{amssymb}

\usepackage{graphicx}
\usepackage[export]{adjustbox}
\usepackage[most]{tcolorbox}

\usepackage{subcaption}
\usepackage{cleveref}
\usepackage{bm}
\usepackage[bbgreekl]{mathbbol}
\DeclareMathSymbol\bbDelta  \mathord{bbold}{"01}

\newcommand{\preprint}[1]{\begin{table}[t]  
    \begin{flushright}      
             {#1}        
             \end{flushright}                 
             \end{table}}  
\renewcommand{\title}[1]{\vbox{\center\LARGE{#1}}\vspace{5mm}}
\renewcommand{\author}[1]{\vbox{\center#1}\vspace{5mm}}
\newcommand{\address}[1]{\vbox{\center\em#1}}

\numberwithin{equation}{section}
\newcommand{\be}{\begin{equation}}
\newcommand{\ee}{\end{equation}}
\newcommand{\bea}{\begin{eqnarray}}
\newcommand{\eea}{\end{eqnarray}}

\Crefname{equation}{Eq.}{Eqs.}

\ifdefined \Ref
\renewcommand{\Ref}[1]{Ref.~\cite{#1}}
\else
\newcommand{\Ref}[1]{Ref.~\cite{#1}}
\fi
\newcommand{\fig}[1]{Fig.~\ref{#1}}

\newcommand{\Eq}[1]{Eq.~\eqref{#1}}

\newcommand{\Sec}[1]{Sec.~\ref{#1}}

\hypersetup{citecolor=blue,linkcolor=blue,urlcolor=blue}

\def\nn{\nonumber}

\newcommand{\calM}{\mathcal{M}}

\newcommand{\calO}{\mathcal{O}}

\newcommand{\probe}{(0)}

\newcommand{\citeLIGO}{LIGOScientific:2016aoc,LIGOScientific:2017vwq}
\newcommand{\citeNewDetectors}{Punturo:2010zz,LISA:2017pwj,Reitze:2019iox}

\newcommand{\citeNRGR}{Goldberger:2004jt}

\usetikzlibrary{snakes}
\usetikzlibrary{decorations}
\usetikzlibrary{trees}
\usetikzlibrary{decorations.pathmorphing}
\usetikzlibrary{decorations.markings}
\usetikzlibrary{external}
\usetikzlibrary{intersections}
\usetikzlibrary{shapes,arrows}
\usetikzlibrary{arrows.meta}
\usetikzlibrary{calc}
\usetikzlibrary{shapes.misc}
\usetikzlibrary{decorations.text}
\usetikzlibrary{backgrounds}
\usetikzlibrary{fadings}
\usetikzlibrary{tikzmark,calc,arrows,shapes,decorations.pathreplacing}

\tikzset{
	graviton/.style={decorate,line width=0.1mm, decoration={snake,amplitude=.5mm, segment length=2mm}},
	photon/.style={decorate, decoration={snake}, draw=red},
	scalar/.style={postaction={decorate},
	},
	massive/.style={postaction={decorate},
		line width=0.75mm,
	},
	massless/.style={postaction={decorate},
	},
  	masslessOverlap/.style={
		decoration={show path construction, lineto code={
			\draw[white,line width=1.5mm,line cap=round] ($(\tikzinputsegmentfirst)!0.2!(\tikzinputsegmentlast)$) to ($(\tikzinputsegmentfirst)!0.8!(\tikzinputsegmentlast)$);
			\draw (\tikzinputsegmentfirst) to (\tikzinputsegmentlast);
		}},
	decorate
	},
	masslessWithDot/.style={postaction={decorate},
		decoration={
			markings,
			mark=at position 0.5 with {\fill circle (2pt);}}
	},
	massiveWithDot/.style={postaction={decorate},
		line width=0.5mm,
		decoration={
			markings,
			mark=at position 0.5 with {\fill circle (2pt);}}
	},
	massiveLinWithDot/.style={postaction={decorate},
		double,
		thick,
		fill=white,
		decoration={
			markings,
			mark=at position 0.5 with {\fill[black] circle (2pt);}}
	},	
	massiveWithArrow/.style={postaction={decorate},
		line width=0.75mm,
		decoration={
			markings,
			mark=at position 0.5 with {\arrow{latex}}}
	},
	massiveWithArrow2/.style={postaction={decorate},
		line width=0.75mm,
		decoration={
			markings,
			mark=at position 0.9 with {\arrow{latex}}}
	},
	massiveLin/.style={postaction={decorate},
		double,
		thick,
		fill=white
	},
	massivePhi/.style={postaction={decorate},
		line width=0.75mm,
		dashed
	},
	masslessPhi/.style={postaction={decorate},
		dashed
	},
	unitaryCut/.style={postaction={draw,densely dashed,blue,thin},
		line width = 0.2cm,white
	},
	gluon/.style={decorate, draw=magenta,
		decoration={coil,amplitude=4pt, segment length=5pt}},
	partial ellipse/.style args={#1:#2:#3}{
		insert path={+ (#1:#3) arc (#1:#2:#3)}
	},
	cross/.style={cross out, draw=black, minimum size=2*(#1-\pgflinewidth), inner sep=0pt, outer sep=0pt},
	branchCut/.style={postaction={decorate},
		snake=zigzag,
		decoration = {snake=zigzag,segment length = 2mm, amplitude = 2mm}	
	},
	cross/.default={1pt},
	massiveWithResidue/.style={postaction={decorate},
		decoration={
			markings,
			mark=at position 0.5 with {\draw node[cross,red] {};}}
	},
	massiveLinWithResidue/.style={postaction={decorate},
	double,
	thick,
	fill=white,
	decoration={
	markings,
	mark=at position 0.5 with {\draw node[cross=4pt,red] {};}}	
	}
}

\colorlet{mred}{black!30!red}
\colorlet{mgreen}{black!30!green}
\colorlet{mblue}{black!30!blue}
\colorlet{morange}{blue!70!red}

\newcommand{\threeptPPh}{
\begin{tikzpicture}[scale=1]
		\coordinate (e1) at (-1,0.);
		\coordinate (e2) at (0.,1);
		\coordinate (e3) at (1,0);
		
		\coordinate (v1) at (0,0);
		
		\draw[massiveLin] (e1) -- (v1) -- (e3) ;
		\draw[graviton](v1) -- (e2) ;

        \node[left]  at (e1) {$\phi_2$};
        \node[above] at (e2) {$h$};
        \node[right] at (e3) {$\phi_2$};
\end{tikzpicture}}

\newcommand{\threeptpph}{
\begin{tikzpicture}[scale=1]
		\coordinate (e1) at (-1,0.);
		\coordinate (e2) at (0.,1);
		\coordinate (e3) at (1,0);
		
		\coordinate (v1) at (0,0);
		
		\draw[massive] (e1) -- (v1) -- (e3) ;
		\draw[graviton](v1) -- (e2) ;

        \node[left]  at (e1) {$\phi_1$};
        \node[above] at (e2) {$h$};
        \node[right] at (e3) {$\phi_1$};
\end{tikzpicture}}

\newcommand{\threeptcch}{
\begin{tikzpicture}[scale=1]
		\coordinate (e1) at (-1,0.);
		\coordinate (e2) at (0.,1);
		\coordinate (e3) at (1,0);
		
		\coordinate (v1) at (0,0);
		
		\draw[masslessPhi] (e1) -- (v1) -- (e3) ;
		\draw[graviton](v1) -- (e2) ;

        \node[left]  at (e1) {$\psi$};
        \node[above] at (e2) {$h$};
        \node[right] at (e3) {$\psi$};
\end{tikzpicture}}

\newcommand{\threeptppc}{
\begin{tikzpicture}[scale=1]
		\coordinate (e1) at (-1,0.);
		\coordinate (e2) at (0.,1);
		\coordinate (e3) at (1,0);
		
		\coordinate (v1) at (0,0);
		
		\draw[massive] (e1) -- (v1) -- (e3) ;
		\draw[masslessPhi](v1) -- (e2) ;

        \node[left]  at (e1) {$\phi_1$};
        \node[above] at (e2) {$\psi$};
        \node[right] at (e3) {$\phi_1$};
\end{tikzpicture}}

\newcommand{\threepthhh}{
\begin{tikzpicture}[scale=1]
		\coordinate (e1) at (-1,0.);
		\coordinate (e2) at (0.,1);
		\coordinate (e3) at (1,0);
		
		\coordinate (v1) at (0,0);
		
		\draw[graviton] (e1) -- (v1) -- (e3) ;
		\draw[graviton](v1) -- (e2) ;

        \node[left]  at (e1) {$h$};
        \node[above] at (e2) {$h$};
        \node[right] at (e3) {$h$};
\end{tikzpicture}}

\newcommand{\twoPMeg}{
\begin{tikzpicture}[scale=1]
		\coordinate (e1) at (-1,0.);
		\coordinate (e2) at (-1,1);
		\coordinate (e3) at (1,1);
		\coordinate (e4) at (1,0.);
		
		\coordinate (v0) at (0,0.5);
		\coordinate (v1) at (-.5,0);
		\coordinate (v2) at (.5,0);
		\coordinate (v3) at (0,1);
		
		\draw[graviton] (v0) -- (v3) ;
		\draw[massiveLin](e2) -- (e3) ;
		\draw[massive](e1) -- (e4) ;
		\draw[masslessPhi](v1) -- (v0) -- (v2) ;

        \node[below left]  at (e1) {$\phi_1$};
        \node[above left] at (e2) {$\phi_2$};
        \node[above right] at (e3) {$\phi_2$};
        \node[below right]  at (e4) {$\phi_1$};
\end{tikzpicture}}

\newcommand{\threePMeg}{
\begin{tikzpicture}[scale=1]
		\coordinate (e1) at (-1,0.);
		\coordinate (e2) at (-1,1);
		\coordinate (e3) at (1,1);
		\coordinate (e4) at (1,0.);
		
		\coordinate (v01) at (-.5,0.5);
		\coordinate (v02) at (.5,0.5);
		\coordinate (v1) at (-.5,0);
		\coordinate (v2) at (.5,0);
		\coordinate (v31) at (-.5,1);
		\coordinate (v32) at (0.5,1);
		
		\draw[graviton] (v01) -- (v31) ;
		\draw[graviton] (v02) -- (v32) ;
		\draw[massiveLin](e2) -- (e3) ;
		\draw[massive](e1) -- (e4) ;
		\draw[masslessPhi](v1) -- (v01) -- (v02) -- (v2) ;

        \node[below left]  at (e1) {$\phi_1$};
        \node[above left] at (e2) {$\phi_2$};
        \node[above right] at (e3) {$\phi_2$};
        \node[below right]  at (e4) {$\phi_1$};
\end{tikzpicture}}

\newcommand{\fourPMeg}{
\begin{tikzpicture}[scale=1]
		\coordinate (e1) at (-1,0.);
		\coordinate (e2) at (-1,1);
		\coordinate (e3) at (1,1);
		\coordinate (e4) at (1,0.);
		
		\coordinate (v01) at (-.5,0.5);
		\coordinate (v03) at (0,0.5);
		\coordinate (v02) at (.5,0.5);
		\coordinate (v1) at (-.5,0);
		\coordinate (v2) at (.5,0);
		\coordinate (v31) at (-.5,1);
		\coordinate (v32) at (0.5,1);
		\coordinate (v33) at (0,1);
		
		\draw[graviton] (v01) -- (v31) ;
		\draw[graviton] (v02) -- (v32) ;
		\draw[graviton] (v03) -- (v33) ;
		\draw[massiveLin](e2) -- (e3) ;
		\draw[massive](e1) -- (e4) ;
		\draw[masslessPhi](v1) -- (v01) -- (v02) -- (v2) ;

        \node[below left]  at (e1) {$\phi_1$};
        \node[above left] at (e2) {$\phi_2$};
        \node[above right] at (e3) {$\phi_2$};
        \node[below right]  at (e4) {$\phi_1$};
\end{tikzpicture}}

\newcommand{\ctTree}{
\begin{tikzpicture}[scale=1]
		\coordinate (e1) at (-1,-1);
		\coordinate (e2) at (-1,1);
		\coordinate (e3) at (1,1);
		\coordinate (e4) at (1,-1);
		
		\coordinate (v1) at (0,0);
		
		\draw[massiveLin] (e2) -- (v1) -- (e3) ;
		\draw[masslessPhi](e1) -- (v1) -- (e4) ;

        \path[draw=black,line width=1.5pt, fill=white] (v1) circle[radius=0.25];
        \path[draw=black, line width=1.5pt] (v1) -- +(0.17,0.17);
        \path[draw=black, line width=1.5pt] (v1) -- +(-0.17,-0.17);
        \path[draw=black, line width=1.5pt] (v1) -- +(0.17,-0.17);
        \path[draw=black, line width=1.5pt] (v1) -- +(-0.17,0.17);
        
        \node[below left]  at (e1) {$\psi(\ell_1)$};
        \node[above left]  at (e2) {$\phi_2(-m_2 u_2)$};
        \node[above right] at (e3) {$\phi_2(m_2 u_2)$};
        \node[below right] at (e4) {$\psi(\ell_4)$};
\end{tikzpicture}}

\newcommand{\ctOneLoopV}{
\begin{tikzpicture}[scale=1]
		\coordinate (e1) at (-1,-1);
		\coordinate (e2) at (-1,1);
		\coordinate (e3) at (1,1);
		\coordinate (e4) at (1,-1);
		
		\coordinate (v1) at (0,.20);
        \coordinate (v2) at (-.6,-1);
        \coordinate (v3) at (.6,-1);
		
		\draw[massiveLin] (e2) -- (v1) -- (e3) ;
        \draw[massive] (e1) -- (e4) ;
		\draw[masslessPhi](v2) -- (v1) -- (v3) ;

        \path[draw=black,line width=1.5pt, fill=white] (v1) circle[radius=0.25];
        \path[draw=black, line width=1.5pt] (v1) -- +(0.17,0.17);
        \path[draw=black, line width=1.5pt] (v1) -- +(-0.17,-0.17);
        \path[draw=black, line width=1.5pt] (v1) -- +(0.17,-0.17);
        \path[draw=black, line width=1.5pt] (v1) -- +(-0.17,0.17);

        \node[below left]  at (e1) {$\phi_1(-p_1)$};
        \node[above left]  at (e2) {$\phi_2(-p_2)$};
        \node[above right] at (e3) {$\phi_2(p_2-q)$};
        \node[below right] at (e4) {$\phi_1(p_1+q)$};
\end{tikzpicture}}

\newcommand{\FivePtTree}{
\begin{tikzpicture}[scale=1]
		\coordinate (e1) at (-1,-1);
		\coordinate (e2) at (-1,1);
		\coordinate (e3) at (1,1);
		\coordinate (e4) at (1,-1);
        \coordinate (e5) at (1.25,0);
		
		\coordinate (v1) at (0,0);
		
		\draw[massiveLin] (e2) -- (v1) -- (e3) ;
        \draw[massive] (e1) -- (v1) -- (e4) ;
		\draw[masslessPhi](v1) -- (e5) ;

        \draw[fill=lightgray, opacity=1] (v1) ellipse (0.45 and 0.45) node {};

        \node[below left]  at (e1) {$\phi_1(-p_1)$};
        \node[above left]  at (e2) {$\phi_2(-p_2)$};
        \node[above right] at (e3) {$\phi_2(p_2+\ell_2)$};
        \node[below right] at (e4) {$\phi_1(p_1+\ell_1)$};
        \node[right]       at (e5) {$\psi(k)$};
\end{tikzpicture}}

\newcommand{\treet}{
\begin{tikzpicture}[scale=1.3]
		\coordinate (e1) at (-0.5,0.);
		\coordinate (e2) at (.5,0.);
		\coordinate (e3) at (.5,-1.);
		\coordinate (e4) at (-0.5,-1.);	
		
		\coordinate (v1) at (0,0);
		\coordinate (v2) at (0,-1.);
		
		\draw[massive] (e1) -- (v1) -- (e2) ;
		\draw[massive] (e3) -- (v2) -- (e4) ;
		
		\draw[graviton] (v1) -- (v2) ;

        \node[below left]  at (e4) {$p_1$};
        \node[above left]  at (e1) {$p_2$};
        \node[above right] at (e2) {$p_3$};
        \node[below right] at (e3) {$p_4$};
\end{tikzpicture}}

\newcommand{\zeroPMradQ}{
\begin{tikzpicture}[scale=1]
		\coordinate (e1) at (-1,0.);
		\coordinate (e2) at (-1,1);
		\coordinate (e3) at (1,1);
		\coordinate (e4) at (1,0.);
		
		\coordinate (v0) at (0.0,0);
		\coordinate (v1) at (.75,.5);
	
		\draw[massiveLin](e2) -- (e3) ;
		\draw[massive](e1) -- (e4) ;
		\draw[masslessPhi](v1) -- (v0)  ;

        \node[below left]  at (e1) {$\phi_1$};
        \node[above left] at (e2) {$\phi_2$};
        \node[above right] at (e3) {$\phi_2$};
        \node[below right]  at (e4) {$\phi_1$};
\end{tikzpicture}}

\newcommand{\onePMradQEgOne}{
\begin{tikzpicture}[scale=1]
		\coordinate (e1) at (-1,0.);
		\coordinate (e2) at (-1,1);
		\coordinate (e3) at (1,1);
		\coordinate (e4) at (1,0.);
		
		\coordinate (v0) at (0,0);
		\coordinate (v1) at (.75,.5);
        \coordinate (v2) at (.25,0);
        \coordinate (v3) at (0,1);
	
		\draw[massiveLin](e2) -- (e3) ;
		\draw[massive](e1) -- (e4) ;
		\draw[masslessPhi](v1) -- (v2)  ;
        \draw[graviton](v0) -- (v3)  ;

        \node[below left]  at (e1) {$\phi_1$};
        \node[above left] at (e2) {$\phi_2$};
        \node[above right] at (e3) {$\phi_2$};
        \node[below right]  at (e4) {$\phi_1$};
\end{tikzpicture}}

\newcommand{\onePMradQEgTwo}{
\begin{tikzpicture}[scale=1]
		\coordinate (e1) at (-1,0.);
		\coordinate (e2) at (-1,1);
		\coordinate (e3) at (1,1);
		\coordinate (e4) at (1,0.);
		
		\coordinate (v0) at (0,0);
		\coordinate (v1) at (.75,.5);
        \coordinate (v2) at (0,.5);
        \coordinate (v3) at (0,1);
	
		\draw[massiveLin](e2) -- (e3) ;
		\draw[massive](e1) -- (e4) ;
		\draw[masslessPhi] (v0)-- (v2) -- (v1)  ;
        \draw[graviton](v2) -- (v3)  ;

        \node[below left]  at (e1) {$\phi_1$};
        \node[above left] at (e2) {$\phi_2$};
        \node[above right] at (e3) {$\phi_2$};
        \node[below right]  at (e4) {$\phi_1$};
\end{tikzpicture}}

\begin{document}

\unitlength = .8mm

\begin{titlepage}
\begin{center}

\hfill \\
\hfill \\
\vskip 1cm

\title{Comparison of post-Minkowskian and self-force expansions: Scattering in a scalar charge toy model}

\author{Leor Barack,$^1$ Zvi Bern,$^2$ Enrico Herrmann,$^2$ Oliver Long,$^{3,1}$ \\
Julio Parra-Martinez,$^4$ Radu Roiban,$^5$ Michael S. Ruf,$^2$ Chia-Hsien Shen,$^6$\\
Mikhail P. Solon,$^2$ Fei Teng,$^5$ and Mao Zeng$^7$}

\address{
${}^1$Mathematical Sciences, University of Southampton, Southampton SO17 1BJ, United Kingdom \\
${}^2$Mani L. Bhaumik Institute for Theoretical Physics,
University of California at Los Angeles,
Los Angeles, CA 90095, USA
\\
${}^3$Max Planck Institute for Gravitational Physics (Albert Einstein Institute), Am M\"uhlenberg 1, Potsdam 14476, Germany
\\
${}^4$Walter Burke Institute for Theoretical Physics,
    California Institute of Technology, Pasadena, CA 91125, USA
\\
${}^5$Institute for Gravitation and the Cosmos,
Pennsylvania State University,
University Park, PA 16802, USA
\\
${}^6$Department of Physics, University of California at San Diego, 9500 Gilman Drive, La Jolla, CA 92093-0319, USA
\\
${}^7$Higgs Centre for Theoretical Physics, University of Edinburgh, James Clerk Maxwell Building, Peter Guthrie Tait Road, Edinburgh, EH9 3FD, United Kingdom
}
\end{center}

\abstract{We compare numerical self-force results and analytical fourth-order post-Minkowskian (PM) calculations for hyperbolic-type scattering of a point-like particle carrying a scalar charge $Q$ off a Schwarzschild black hole, showing a remarkably good agreement. Specifically, we numerically compute the scattering angle including the full $\mathcal{O}(Q^2)$ scalar-field self-force term (but ignoring the gravitational self-force), and compare with analytical expressions obtained in a PM framework using scattering-amplitude methods.  This example provides a nontrivial, high-precision test of both calculation methods, and illustrates the complementarity of the two approaches in the context of the program to provide high-precision models of gravitational two-body dynamics. Our PM calculation is carried out through 4PM order, i.e., including all terms through $\mathcal{O}(Q^2 G^3)$. At the fourth post-Minkowskian order the point-particle description involves two a-priori undetermined coefficients, due to contributions from tidal effects in the model under consideration.  These coefficients are chosen to align the post-Minkowskian results with the self-force ones. 
}

\vfill

\restoregeometry

\end{titlepage}

\eject

\begingroup
    \hypersetup{linkcolor=black}
    \tableofcontents
\endgroup

\newpage

\section{Introduction}
\label{sec:intro}

The landmark detection of gravitational waves by the LIGO and Virgo collaborations~\cite{\citeLIGO} has opened an era of scientific exploration promising new and unexpected discoveries in astronomy, cosmology, and particle physics.  To make full use of the anticipated vast improvements in the precision of new planned detectors~\cite{\citeNewDetectors} requires a commensurate improvement in theoretical waveform predictions.  Achieving this will require input from multiple complementary approaches, including numerical relativity~\cite{Pretorius:2005gq, baumgarte_shapiro_2010, Damour:2014afa, Hopper:2022rwo}, the self-force (SF) approach~\cite{Mino:1996nk,Quinn:1996am, Poisson:2011nh, Barack:2018yvs}, effective field theory~\cite{\citeNRGR}, as well as weak-field perturbative treatments based on the post-Newtonian (PN)~\cite{Droste:1916, Droste:1917, Einstein:1938yz,Ohta:1973je} or post-Minkowskian (PM)~\cite{Bertotti:1956pxu, Kerr:1959zlt, Bertotti:1960wuq, Westpfahl:1979gu, Portilla:1980uz, Bel:1981be} frameworks. These inputs can then be combined into accurate waveform models such as those based on the effective-one-body (EOB) approach~\cite{Buonanno:1998gg,Buonanno:2000ef}.

\begin{figure}[t]
\centering
\includegraphics[width=0.9\linewidth]{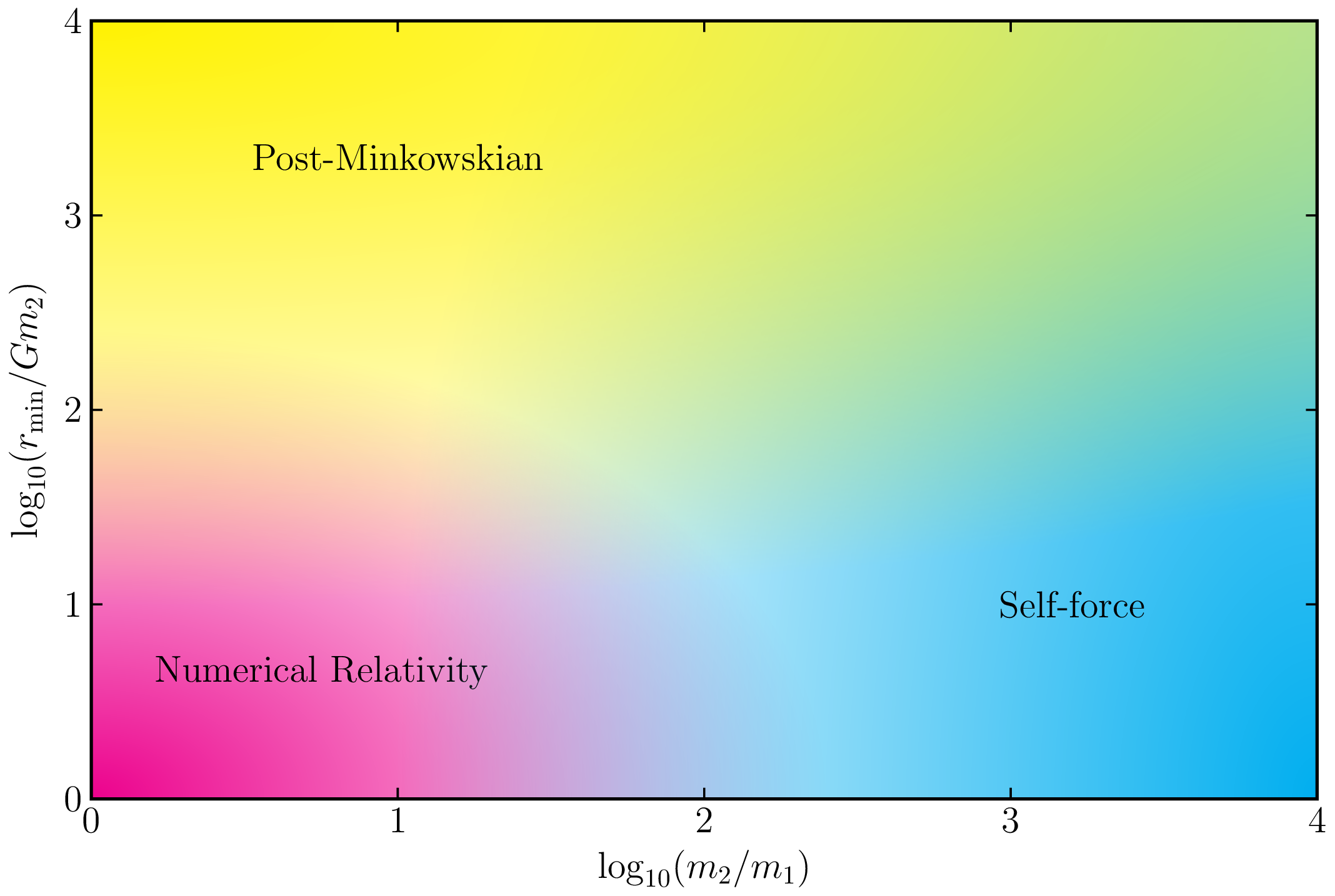}
\caption{ Domains of the two-body scattering problem in GR. The diagram schematically shows the main calculation techniques in their respective natural domains.  Overlap regions provide arenas for comparison and exchange of complementary information. In this paper we explore the overlap region at the top right of the plot, i.e.\ the domain of weak-field scattering at large mass-ratio.}
\label{TwoBodyDiagram}
\end{figure}

Recent years have witnessed a renewed interest in the {\it scattering} regime of the binary black-hole problem, a process that provides a clean theoretical probe of the two-body dynamics. In the scattering setup, the initial and final states are at large separations, where the background spacetime is asymptotically Minkowskian, allowing an unambiguous definition of physical observables such as the momentum impulse and the scattering angle. This, in turn, facilitates the comparison of results obtained using different approaches; see Fig.\ \ref{TwoBodyDiagram}. In this paper, we carry out a proof-of-principle high-precision comparison of the post-Minkowskian (PM) and self-force (SF) approaches in their overlap region, in the context of a scalar-charge toy model~\cite{Quinn:2000wa, Gralla:2021qaf}.  We cross-confirm the consistency of results and illustrate how each approach informs and complements the other.

The PM framework is a natural perturbative approach to the scattering problem at large minimum separation. In this framework, the two black holes are treated as point particles in a scattering process on a fixed flat background. The gravitational interaction between the two particles is accounted for order by order in Newton's constant $G$, with the velocity kept arbitrary in a format that respects special relativity. In this framework, finite size effects, such as tidal deformability and horizon dissipation can be included perturbatively in terms of higher dimension interactions in an effective field theory. 

A new approach within the PM framework is the scattering-amplitudes method, which leverages enormous advances in computing and understanding scattering amplitudes.   These advances include generalized unitarity~\cite{Bern:1994zx, Bern:1994cg,Bern:1995db, Bern:1997sc, Britto:2004nc}, the double-copy construction~\cite{Kawai:1985xq, Bern:2008qj, Bern:2010ue, Bern:2019prr} and powerful integration methods~\cite{Chetyrkin:1981qh, Laporta:2000dsw, Kotikov:1990kg, Bern:1993kr, Remiddi:1997ny, Gehrmann:1999as}.  They have enabled explicit gauge and gravity calculations at remarkably high orders of perturbation theory (see e.g.~\cite{Bern:1994zx, Berger:2008sj, Berger:2010zx,  Bern:2012uf, Bern:2012uc, Bern:2018jmv}). The double copy expresses gravitational scattering amplitudes in terms of simpler corresponding gauge-theory amplitudes, while generalized unitarity gives a means for building loop amplitudes from simpler tree amplitudes.  More recently they have been applied in the context of gravitational-wave physics~\cite{Cheung:2018wkq, Kosower:2018adc, Bern:2019nnu, Bern:2019crd, Parra-Martinez:2020dzs, Bern:2021dqo, Bern:2021yeh}, where they led to rapid progress in the development of PM theory for binary systems in General Relativity (GR). In the classical limit, the scattering amplitudes are directly useful objects, because they are simply related to the Hamilton-Jacobi radial action~\cite{Bern:2021dqo}.

A basic premise of the quantum scattering-amplitude approach to gravity is that in the weak-field regime, gravitational forces are mediated by massless spin-2 particles~\cite{Feynman:1996kb}, which are identified as the fluctuations of the space-time metric around a flat Minkowski background.
This perspective realizes the PN and PM frameworks as quantum field theory perturbation theory around flat space.
Remarkably, even in the presence of their interactions arising from matter-coupled Einstein-Hilbert action, the spin-2 particles can be thought of as two copies of spin-1 particles~\cite{Kawai:1985xq, Bern:2008qj, Bern:2010ue, Bern:2019prr}, emphasizing the close connection of gravitational and gauge theories. 
Although the starting point is rather different from the classic one based on Einstein's equations, the final physical predictions are identical.

The SF approach \cite{Barack:2018yvs} is also perturbative, but it does not rely on a weak-field approximation. Instead, it is based on an expansion in the mass ratio of the binary, assumed small.  At leading order, the lighter object traces a timelike geodesic in the exact spacetime geometry of the larger object (e.g., Kerr geometry). One then systematically incorporates post-geodesic terms order by order in the mass ratio. This approach is ``exact'' in the strength of the gravitational interaction and thus applicable in the strong-field regime, as long as the mass ratio is small. The self-force program has been primarily motivated by the prospect of observing gravitational waves from the inspiral of compact objects into a massive black hole with planned observatories in space. For that reason, much of the focus has been on bound-orbit configurations. There has been significant progress over the past few years, culminating in the recent milestone (numerical) calculation of a full quasi-circular inspiral orbit and its emitted gravitational waveform, which necessitated the inclusion of second-order  gravitational SF terms~\cite{Wardell:2021fyy}. SF calculations have also been useful in setting accurate strong-field benchmarks to inform other modeling approaches, primarily PN methods and EOB. In particular, they have been utilized to calibrate a priori unknown terms in the EOB potentials \cite{Barack:2010ny, Nagar:2022fep}, to resolve initial ambiguities in PN results \cite{Detweiler:2008ft, Barack:2011ed, Bini:2013zaa, vandeMeent:2016hel}, and are key to  the ``Tutti-Frutti'' approach \cite{Bini:2019nra,Bini:2020wpo, Antonelli:2020aeb, Antonelli:2020ybz, Khalil:2021fpm}. Very recent work has started applying the SF method to scattering orbits~\cite{Hopper:2017qus, Hopper:2017iyq, Gralla:2021qaf, Gralla:2021eoi, Long:2021ufh, Barack:2022pde}.

The SF and PM approaches are complementary, see Fig.~\ref{TwoBodyDiagram}: for strong-field scattering, the PM approximation breaks down, while for weak-field scattering numerical SF calculations become less tractable, for a variety of technical reasons to be explained later.   Nevertheless, one can make detailed comparisons in an intermediate regime where both approaches remain controllable.  As noted by Damour, simple mass dependence in the PM approach implies simple rules for the overlap of PM and SF orders~\cite{Damour:2019lcq}.  The comparisons carried out here are meant as a proof-of-principle demonstration. We will discuss planned follow-on improvements, which include, on the SF side, an improved methodology for the numerical integration of the relevant field equations; and, on the PM side, the implementation of resummation techniques \`{a} la EOB, where they have been shown to remarkably improve the agreement with numerical relativity computations when 4PM contributions are incorporated~\cite{Khalil:2022ylj, Damour:2022ybd}.

For this initial study, we adopt the scalar-charge toy model described in detail in Ref.~\cite{Barack:2022pde}, and we also use the time-domain computational infrastructure developed in that work for numerically calculating the scattering angle including SF effects from the scalar-field back-reaction. Here we introduce various code optimizations and several new tricks to enable evaluation in the PM domain, which is computationally more demanding. Our analytical post-Minkowskian calculations follow those of Refs.~\cite{Bern:2019nnu, Bern:2019crd, Bern:2021dqo, Bern:2021yeh} for the conservative part and those of Refs.~\cite{Damour:2020tta, Herrmann:2021lqe, Herrmann:2021tct, Bini:2021gat,
Manohar:2022dea} for the radiative contributions, except that instead of pure gravity, an additional massless scalar field is incorporated, amounting to the evaluation of additional diagrams in the same setup.  Since the foundations of our method are described elsewhere, we will review them only briefly here, instead focusing on new features of the current calculation. 

Here we are interested in demonstrating the ability to carry out precision comparisons between self-force and post-Minkowskian calculations in an overlap region where both approaches are valid. To carry out a comparison we analyze black hole scattering in the region where one object is much lighter than the other, the scalar charge is ``small'', and the minimum separation distance is sufficiently large. The smallness of the charge is expressed as a condition on a certain dimensionless combination $q_s$ of the charge and the masses [see Eq.~(\ref{eq:qs_def}) below], which guarantees that the scattering orbit of the small object deviates only slightly from a geodesic in the background spacetime of the larger object, assumed to be a Schwarzschild black hole. Geodesic motion is recovered in the limit $q_s\to 0$, and in our calculation of the scattering angle we keep only terms through $q_s)$, i.e.\ we account for (only) the first-order back-reaction effect from the scalar-field SF. The gravitational SF is altogether neglected.     
To align the PM calculation with the SF one, we similarly truncate the PM contributions, keeping only terms up to $q_s)$, and dropping all gravitational SF terms. Note that, in this approximation, the mass dependence of the PM expansion terms is trivial at all orders: the $N$th-order PM term of the self-force correction to the scattering angle depends on the masses through a factor $q_s m_2^N$, where $m_2$ is the mass of the large black hole.

The leading-order scalar-field SF correction to the geodesic scattering angle, which is a 2PM effect, has been derived analytically by Gralla and Lobo in \cite{Gralla:2021qaf}, using SF methods. Here we extend this through to 4PM order, using effective-field methods. Our calculation includes both conservative and dissipative terms, and it is complete through 3PM order.

The point-particle framework used in both the PN and PM expansions 
must be extended by the inclusion of tidal operators to describe all details of compact objects of finite size.  
This is an ubiquitous feature of effective field theory, see e.g.~Sec.~2.6 of the review~\cite{Goldberger:2007hy}, whose action includes operators of increasingly higher dimension as one probes distances close to the size of the object. 
Quite generally, at sufficiently high order in perturbation theory, point-particle and finite-size effects are not cleanly separated. This is familiar in quantum field theory and leads to the need for renormalization. Power counting dictates the order at which this occurs and the general structure of the required (counter)terms.
The classical limit delays the appearance of tidal operators compared to the quantum case; for example, together with diffeomorphism invariance, it implies that in the classical limit of gravitational theories the first 
tidal (two-curvature) operator is first required at 6PM (five-loop) order while in the quantum theory they are already required at one-loop.
The power counting in the scalar-charge model is different from that of gravity because, unlike the graviton field, the scalar field does not need to be dressed with additional derivatives to preserve gauge invariance.  This difference implies that, already at 4PM (three-loop) order, two-massive-two-massless-scalar operators are required for the model to be well-defined.
At this order a short-distance divergence appears in the classical regime, indicating that the objects cannot be assumed to be point-like. 
Tidal operators are needed to absorb this divergence via the standard renormalization process. The value of the corresponding renormalized coupling---the scalar Love numbers---are among the parameters characterizing compact objects with scalar charge.\footnote{Such divergences also appear if one computed the scattering angle of two delta-function sources by solving perturbatively Einstein's equations and it can be cured by using instead an extended source. From this perspective, the Love numbers encode characteristics the renormalized source. See e.g. Sec.~8 of Ref.~\cite{Bern:2019prr} 
for a simple related illustration, in which the total mass of the point-like source is renormalized. The Love numbers may be determined by matching an observable with the response of the finite-size object to massless-scalar perturbations in the scalar-charge model and gravitational perturbations in GR. } 

In contrast, in the SF formalism, the system is {\em exactly} described by a finite-sized central black hole with a point-particle companion at leading order, with finite-size effects of the secondary being incorporated at higher orders in the mass ratio.  The net effect of the tidal operators is to introduce additional parameters that can be fixed by matching to calculations in black hole perturbation theory~\cite{Fang:2005qq, Damour:2009vw, Binnington:2009bb, Kol:2011vg, Landry:2015zfa, LeTiec:2020bos, Chia:2020yla, Hui:2020xxx, Charalambous:2021mea, Hui:2021vcv, Ivanov:2022qqt, Ivanov:2022hlo}. Matching is a standard procedure in effective field theory and leads to well defined results. Here we will not carry out such a calculation but instead will use the comparison to the corresponding SF calculation in an attempt to (numerically) determine the values of these parameters.  While this reduces our ability to carry out a precision comparison of the two approaches at 4PM order, it also illustrates how one approach can aid the other to resolve ambiguities.

This paper is organized as follows.  In Sec.~\ref{sec:ScalarSF} we briefly review the foundations of the SF approach as applied to the scattering problem, and in \Sec{sec:PM} we give a similar overview of the amplitudes-inspired approach of Refs.~\cite{Bern:2021dqo, Bern:2021yeh, Manohar:2022dea} used here to carry out the PM calculation through 4PM order. \Sec{sec:NumericalSF} reviews the numerical method developed in Refs.~\cite{Long:2021ufh, Barack:2022pde} and the various improvements that were introduced to facilitate our comparison with PM results.  In \Sec{sec:PM_results} we present analytic results for the conservative and dissipative scattering angle through 4PM order, and we also explain the appearance of scalar tidal operators. The results from the two approaches are then compared in the overlap region in \Sec{sec:Comparison}. \Sec{sec:Outlook} contains a summary and a road map for future perusal of our program.  An appendix clarifies that the effect of center-of-mass boost from the net loss of linear momentum in radiated scalar waves is negligible within our working assumptions. 

We use units where $ c=1$. The gravitational constant $G$ is usually set to 1, but we restore factors of $G$ when counting PM interaction orders, for clarity. Similarly, we usually set $\hbar = 1$ but restore it as a counting parameter for the classical limit. We work in mostly-plus metric signature and reserve boldface symbols for spatial components of four-vectors.

\section{Self-force description}
\label{sec:ScalarSF}

\subsection{Overview of self-force approach}

Anticipating a readership of mixed expertise, we start with a brief general overview of the self-force approach to the two-body dynamics in GR. The topic has been reviewed thoroughly in recent literature. We can especially refer non-expert readers to \cite{Barack:2018yvs} for an introductory-level exposition. 

SF theory sits at the intersection between two fundamental problems in classical GR: the problem of motion (``how do objects move in a curved spacetime?'') and the two-body problem (``what is the spacetime of a binary system of gravitationally-interacting compact objects?''). Work on both problems goes back a century to the early days of GR, but the advent of gravitational-wave astronomy has given them renewed relevance and an important modern context. The SF program deals with the modeling of the inspiral dynamics in compact-object binaries with extreme mass ratios, and the prediction of the gravitational-wave signature of such systems. In its modern incarnation, the program took off in the 1990s when concrete plans began for the design of a space-based observatory---LISA, the laser interferometer space antenna---that would give access to the milliHertz band of the gravitational-wave spectrum. A prime target for LISA (now scheduled for launch in the mid-2030s) will be the radiative inspiral of compact stars and stellar-mass black holes into massive black holes in galactic centers: so-called EMRIs (extreme mass ratio inspirals). Due to the large mass ratio, the inspiral is slow, with a typical LISA EMRI giving off some $\sim \! 10^5$ gravitational-wave cycles all while in a tight orbit right outside the massive black hole. EMRI signals will thus be excellent probes of black-hole geometry, and of GR theory itself, in its most extreme regime. But, for that same reason, their modeling cannot rely on traditional weak-field approaches such as the PN or PM frameworks. Full numerical-relativity simulations are also unsuitable because the large-scale disparity makes EMRI simulations computationally intractable. 

SF theory provides a natural approach to EMRI modeling. Like PN and PM theories, it is based on a systematic perturbative treatment of the Einstein equations. But instead of perturbing in the strength of the gravitational coupling ($G$) or in the velocity ($v/c$), in SF theory one perturbs in the mass ratio 
\begin{align}
\label{eq:qm_def}
q_m:=m_1/m_2\ll 1,
\end{align}
keeping the strength of gravity and the velocities arbitrary. In the formulation of the SF equations of motion, there is no a priori reduction of field degrees of freedom to particle degrees of freedom. Rather, there is a systematic solution of the Einstein field equations for the EMRI system order by order in $q_m$, using a procedure of matched asymptotic expansions that takes advantage of the disparate scales in the problem. One assumes only the validity of the vacuum Einstein equations, local conservation of energy-momentum, and the existence of certain formal limits that arise in the matched-expansions procedure. The end result is a description of the EMRI dynamics in terms of a point-like particle that moves along some accelerated worldline in the background geometry of the massive black hole. But the notions of a point particle, point-particle worldline, or point-particle momentum  are all {\em derived}  (not assumed) notions in the theory.

The SF description of an EMRI starts with a formal expansion of the full spacetime metric ${\sf g}_{\alpha\beta}$ in powers of $q_m$ about the exact metric $g_{\alpha\beta}$ of the large black hole when in isolation:
\begin{equation}
{\sf g}_{\alpha\beta} = g_{\alpha\beta} + q_m h^{(1)}_{\alpha\beta} + q_m^2 h^{(2)}_{\alpha\beta} + \cdots.
\end{equation}
Here $g_{\alpha\beta}$ is usually taken to be the Kerr metric (or its nonspinning reduction, Schwarzschild), and $h^{(n)}_{\alpha\beta}$ are $q_m$-independent metric perturbations. This description breaks down close to the small body, where its gravitational field can no longer be considered a perturbation. Instead, near the small object, at distances $d\ll m_2$ from it, one introduces a different expansion of the full metric, in powers of $d/m_2$ this time, where the zeroth order is now the exact metric of the {\em small} object, and higher-order terms describe weak tidal-type perturbations from the external geometry associated with the large black hole.  The assumed scale disparity $m_1\ll m_2$ means that there is a buffer zone $m_1\ll d\ll m_2$ where both asymptotic descriptions apply. One proceeds by matching the two expansions in the buffer zone, term by term in $q_m$, at the end taking the limit $q_m\to 0$ where both mass and size of the small object shrink to zero commensurately. This procedure produces a reference worldline in the exact background geometry $g_{\alpha\beta}$, whose acceleration in that geometry is determined by the matching. This is the inspiral trajectory. The matching also determines the perturbation fields $h_{\alpha\beta}^{(n)}$ (and hence the gravitational waveforms), once physical boundary conditions are supplied.\footnote{Of course, we are glossing here many crucial details, including the need to gauge-fix diffeomorphism invariance/covariance so that the field equations manifestly admit a well-posed initial-value formulation, or the need to account for gauge mismatch in the matching procedure.} 

At leading order in $q_m$, the matching tells us that the worldline representing the small object's trajectory is a geodesic of the Kerr geometry. This is a {\em derivation} of the familiar ``test particles move on geodesics'' maxim. The matching also tells us that the perturbation field $h^{(1)}_{\alpha\beta}$  is a (retarded) solution of the linearized Einstein equations with a delta-function source of stress-energy moving along the geodesic. This provides an effective notion of ``point-particle stress-energy'' in GR.  At the next order in $q_m$, the particle's worldline picks up a small, $\calO(q_m)$ acceleration, interpreted as an effect of a back-reaction force---``self-force''---from  $h^{(1)}_{\alpha\beta}$. This SF drives the inspiral motion, sending off orbital energy and angular momentum in gravitational waves. One can continue in this fashion to obtain the second-order perturbation $h^{(2)}_{\alpha\beta}$, which is made of a nonlinear piece of the metric together with a correction to the linear piece due to the worldline's acceleration. From $h^{(2)}_{\alpha\beta}$ one can calculate the $\calO(q_m^2)$ back-reaction to the particle's acceleration, and so on.  

Restricting to the first-order SF effect, as in the rest of this paper, the particle's equation of motion (in the frame of the large black hole) has the form 
\begin{equation}
m_1 u^\beta \nabla_\beta u^\alpha = F^\alpha,
\label{eqn:SFdefinition}
\end{equation}
where $u^\alpha$ is the particle's four-velocity, $\nabla_\alpha$ is a covariant derivative compatible with the (Kerr) background metric $g_{\alpha\beta}$, and $F^\alpha(\propto q_m^2)$ is the first-order SF. This $F^\alpha$ (a function along the worldline) represents the back-reaction force from the (retarded) linear perturbation $h^{(1)}_{\alpha\beta}$. More precisely, it is constructed from the gradient of a certain ``regular'' piece of $h^{(1)}_{\alpha\beta}$, denoted $h^{(1)R}_{\alpha\beta}$, evaluated at the particle's location. The identification of the correct regular piece $h^{(1)R}_{\alpha\beta}$ comes out automatically out of the matched asymptotic expansions procedure, and does not involve any ad-hoc regularization. 

A great deal of effort has gone into the formulation of practical methods for (1) solving the linearized field equations for $h^{(1)}_{\alpha\beta}$ in an EMRI scenario, and (2) extracting the regular piece $h^{(1)R}_{\alpha\beta}$ from that solution. For general EMRI orbits there are no known analytical solutions for $h^{(1)}_{\alpha\beta}$, so this part of the calculation has to be done numerically. However, since the relevant field equations are linear, this can be done with relatively high numerical precision.  To construct the SF $F^\alpha$ from $h^{(1)}_{\alpha\beta}$, many calculations use the {\em mode-sum method}, also to be used in our work. In this method, the perturbation $h^{(1)}_{\alpha\beta}$ is first decomposed into multipole modes $h^{(1)\ell}_{\alpha\beta}$ (spherical-harmonic modes or similar) on spheres around the central black hole, and the extraction of the regular piece is then conveniently done mode by mode, using a formula of the form  
\begin{equation}\label{mode-sum}
F^\alpha=\sum_{\ell=0}^\infty\left(\nabla^{\alpha\beta\gamma}h_{\beta\gamma}^{(1)\ell}-A^\alpha\ell -B^\alpha -C^\alpha/\ell\right),
\end{equation}
Here $\nabla^{\alpha\beta\gamma}$ is a certain derivative operator, and $A^\alpha,B^\alpha$ and $C^\alpha$ are certain ``regularization parameters'', which are known analytically. While the full perturbation $h^{(1)}_{\alpha\beta}$ is singular at the particle, its individual modes $h^{(1)\ell}_{\alpha\beta}$ and their derivatives are finite there, so the mode-sum procedure circumvents having to deal with infinities in the calculation. We stress that there is no ad-hoc regularization involved in the formulation of the mode-sum procedure. Rather, Eq.~(\ref{mode-sum}) is simply a mode-sum recasting of the extraction formula for $h^{(1)R}_{\alpha\beta}$, itself a result of the rigorous matched expansion method. 

A preliminary task in SF calculations for EMRIs is the evaluation of $F^\alpha$ along fixed geodesic orbits of the Kerr geometry. The actual long-term inspiral evolution of the orbit under the SF effect is computed as a second step, usually using methods from multiple-scale perturbation theory, exploiting the two disparate timescales in the problem: the radiation-reaction timescale is much longer than the orbital timescale. The situation is different in the scattering problem to be considered in this paper, where the slow timescale is lost. Here, in order to consider the $\calO(q_m)$ correction to the geodesic scattering angle, it will suffice to integrate the first-order SF $F^\alpha$ along a fixed scattering geodesic, without actually incorporating the acceleration due to the SF; such a correction to the orbit would only contribute to the scattering angle at $\calO(q_m^2)$, which is neglected in our calculation. 

\subsection{Scalar-field self-force}

Historically, the black-hole scattering scenario has not been given much attention within the SF program, due to its limited direct astrophysical relevance. So while there are today powerful tools for computing EMRI orbits in inspiral scenarios, there are to date no calculations of the full gravitational SF on scattering orbits. Work so far has been restricted to numerical energy flux calculations \cite{Hopper:2017qus, Hopper:2017iyq}, and analytical SF calculations at leading PM order \cite{Gralla:2021qaf}. In Ref.\ \cite{Barack:2022pde} two of us started the development of a complete SF computational infrastructure for scattering orbit, working with a scalar-field toy model as a development platform. The scalar-field model has been employed extensively in SF studies, because it allows to tackle many of the challenging aspects of SF calculations in a simpler environment, without having to solve the full set of linearized Einstein equations or having to deal with complications related to  gauge dependence of the perturbation field. At the same time, the scalar-field model shares many technical similarities with the pure-gravity case, which makes it an excellent laboratory for test and development. 

Our scalar-charge model is described in detail in Sec.~VII of Ref.~\cite{Barack:2022pde}. The larger object is taken to be a Schwarzschild black-hole with mass parameter $m_2$. The smaller object (of mass $m_1$) is endowed  with a scalar charge $Q$, such that 
\begin{align}
\label{eq:qs_def}
q_s:= \frac{Q^2}{m_1 m_2}\ll G= 1\, .
\end{align}
The small parameter $q_s$ is analogous to $q_m$ in the pure-gravity problem, and the condition in (\ref{eq:qs_def}) ensures (see below) that the SF exerted by the scalar field causes only a small perturbation to the geodesic motion of the charged particle. 
The charge sources a massless, minimally coupled Klein-Gordon field $\Psi$, according to 
\begin{equation}
\nabla^\alpha\nabla_\alpha \Psi = -4\pi Q \int_{-\infty}^{\infty} \frac{\delta^4(x-x_p(\tau))}{\sqrt{-g(x)}}d\tau\, .
\label{eqn:Scalar3+1}
\end{equation}
Here $g$ is the determinant of the background Schwarzschild metric $g_{\alpha\beta}$, $\nabla_\alpha$ are covariant derivatives compatible with $g_{\alpha\beta}$, and $x_p(\tau)$ describes the timelike worldline of the pointlike charge in the background geometry, parameterized with proper-time $\tau$. In our work, this worldline is taken to be a scattering geodesic of $g_{\alpha\beta}$. The relevant solution of Eq.\ (\ref{eqn:Scalar3+1}) is the one satisfying retarded boundary condition, i.e.~no incoming radiation from past null infinity, and no outgoing radiation from the past event horizon. We call that solution (which is unique) the ``retarded field''. 

In our model, we completely ignore the gravitational effect of $m_1$, and in particular we ignore the gravitational SF acting on the charge. This is formally achieved by taking the limit $m_1\to 0$ at a fixed $q_s$. We only take into account the back-reaction from the scalar field $\Psi$, which exerts a SF given by
\begin{equation}
Q\nabla^\alpha\Psi^R \propto Q^2.
\end{equation}
Here $\Psi^R$ is the ``regular'' piece of the retarded field, analogous to $h_{\alpha\beta}^{(1)R}$, whose construction has been prescribed using matched asymptotic expansions \cite{Quinn:2000wa}. The scalar-field SF can be obtained using a mode-sum formula as in (\ref{mode-sum}), with analytically known regularization parameters \cite{Barack:2001gx}. The charge's equation of motion is then given by 
\begin{equation} \label{eqn:SFProj}
u^\beta \nabla_\beta (m_1 u^\alpha) = Q\nabla^\alpha\Psi^R ,
\end{equation}
where $u^\alpha=dx^\alpha_p/d\tau$ is the four-velocity.

The self-acceleration of the charge's worldline is given by the component of Eq.~(\ref{eqn:SFProj}) orthogonal to $u^\alpha$, while the tangent component describes an exchange of rest mass $m_1$ with scalar-field energy:
\begin{equation} \label{eqn:SF_orth}
u^\beta \nabla_\beta u^\alpha = 
(\delta_{\beta}^{\alpha}+u^\alpha u_\beta)(Q/m_1)\nabla^\beta\Psi^R =:F^\alpha,
\end{equation}
\begin{equation} \label{eqn:SF_tangent}
\frac{dm_1}{d\tau}=-Q u^\alpha \nabla_\alpha\Psi^R.
\end{equation}
Equation (\ref{eqn:SF_tangent}) can be immediately integrated to give $m_1(\tau)= m_1^{(0)}-Q\Psi^R(\tau)$ (where $m_1^{(0)}$ is an integration constant), which describes the variation of $m_1$ along the worldline. There is no net variation in the scattering case, since $\Psi^R(-\infty)=\Psi^R(+\infty)$. Moreover, the effect of evolving mass $m_1$ enters the acceleration in Eq.\ (\ref{eqn:SF_orth}) only at $\calO(Q^4)$, to be neglected in our analysis. We therefore henceforth ignore the variation in $m_1$, and assume $m_1$ is constant. 

We note that the self-acceleration in Eq.\ (\ref{eqn:SF_orth}) is proportional to $Q^2/m_1\ll m_2$, by our assumption (\ref{eq:qs_def}). Therefore, it drives a very small change in the trajectory of the charge over the length and time scales set by the central black hole. The change this self-acceleration causes to the scattering angle (say, at fixed initial velocity and impact parameters; see below) is of $\calO(q_s)$ and hence also very small. For the purpose of deriving this $\calO(q_s)$ change, it is therefore justified to integrate the SF $F^\alpha$ along the baseline geodesic itself, without correcting for the small change. 

\subsection{Self-force correction to the scattering angle}

A description of scattering orbits through $\calO(q_s)$ is given in detail in Ref.~\cite{Barack:2022pde}. Here we review the essential results. We work in the background spacetime of the black-hole of mass $m_2$, where we use standard Schwarzschild coordinates in which the scattering orbit is described by $\big(t_p(\tau),r_p(\tau),\theta_p(\tau),\varphi_p(\tau)\big)$; without loss of generality we set $\tau=0$ at the periastron, i.e.~$u^r\big|_{\tau=0}=dr_p/d\tau\big|_{\tau=0}=0$. Also without loss of generality, we let the orbit lie in the equatorial plane, $\theta_p\equiv \pi/2$, which the symmetry of the setup allows us to do even with the SF effect included. 

Scattering orbits in Schwarzschild spacetime constitute a 2-parameter family. We choose as parameters the magnitude of the 3-velocity at infinity,
\begin{equation}
v := \frac{1}{u^t_p}\sqrt{(u^r)^2+(r_pu^\varphi)^2}\Big|_{\tau\to -\infty} ,
\end{equation}
and the ``impact parameter'',
\begin{equation}
    b:=\lim_{\tau\to -\infty} r_p \sin \left|\varphi_p(\tau)-\varphi_p(\infty)\right|.
\end{equation}             
We also introduce the parameters 
\begin{equation}\label{sigma&J}
    \sigma = (1-v^2)^{-1/2}\quad\quad\text{and}\quad\quad
    J/m_1 = bv\sigma , 
\end{equation}
which, in the geodesic limit $q_s\to 0$, coincide with the conserved (specific) energy and angular momentum of the scattering orbit in the black-hole frame (usually denoted $E$ and $L$ in SF literature).  Note that both $v$ and $b$ are defined from the asymptotic behavior of the orbit at infinity, which will help us identify orbits across the SF-PM descriptions. 

The scattering angle is defined as
\begin{equation}
\label{eqn:ScatteringAngleDef}
\chi(v,b) := \int^\infty_{-\infty} u^\varphi(\tau;v,b) \, d\tau -\pi .
\end{equation}
In the SF description, we split $\chi(v,b)$ into a ``geodesic'' term and a ``SF'' term, in the form 
\begin{equation}
\chi(v,b) = \chi^{\probe}(v,b) + q_s\: \delta\chi(v,b) + \calO(q_s^2),
\end{equation}
where $\chi^{\probe}(v,p)$ represents the scattering angle for a geodesic orbit with the same $v,b$ as the full orbit. (This is a point of subtlety: fixing other parameters may result in a different split between a ``geodesic term'' and a ``SF term''; we insist on fixing $v,b$ in our split, for the reason just mentioned.)  The geodesic term can be written analytically in terms of an elliptic function:
\begin{equation}\label{eqn:ScatteringAngleDef0}
\chi^{\probe} =
4\sqrt{\frac{p}{p-6-2e}}\;\; {\rm El}_1 \left(\frac{1}{2}\arccos(-1/e);-\frac{4e}{p-6-2e}\right) - \pi.
\end{equation}
Here 
${\rm El}_1$ is the incomplete elliptic integral of the first kind,
\begin{equation}\label{eqn:El1}
{\rm El}_1(\varphi;k):=\int_0^{\varphi} (1-k\sin^2 x)^{-1/2}dx,
\end{equation}
and $e$ (``eccentricity'') and $p$ (``semi-latus rectum'') are related to $\sigma$ and $J$ (and thus to $v$ and $b$) via 
\begin{equation}
    \sigma^2=\frac{(p-2)^2-4e^2}{p(p-3-e^2)}, \hskip 1.5 cm 
    (J/m_1)^2=\frac{p^2 m_2^2}{p-3-e^2}.
\end{equation}

An explicit expression for the SF correction term $\delta\chi$, in terms of an integral of the SF along the geodesic scattering orbit, was derived in Ref.~\cite{Barack:2022pde}. It has the form
\begin{equation}\label{deltaphi1_final_method2}
\delta\chi=
 \int_{-\infty}^{\infty} \Big[{\cal G}_E(\tau) F_t(\tau)
+{\cal G}_L(\tau) F_\varphi(\tau)\Big] d\tau ,
\end{equation}
where ${\cal G}_E$ and ${\cal G}_L$ are (complicated, but analytically given) functions of $\tau$ and the geodesic parameters alone, given in Eq.~(110) of Ref.~\cite{Barack:2022pde}. 

Equation (\ref{deltaphi1_final_method2}) describes the full SF effect, including both conservative and dissipative terms: $\delta\chi= \delta\chi^{\rm cons}+\delta\chi^{\rm diss}$. For the sake of our comparison with PM results it is useful to compute these two pieces separately. At $\calO(q_s)$ the split between $\delta\chi^{\rm cons}$ and $\delta\chi^{\rm diss}$ is defined unambiguously, as follows. First, we define the conservative and dissipative pieces of the SF, using 
\begin{equation}
    F_\alpha^{\rm cons}:= \frac{1}{2}\big(F_\alpha^{\rm ret}+F_\alpha^{\rm adv}\big),\hskip 1.8 cm 
    F_\alpha^{\rm diss}:= \frac{1}{2}\big(F_\alpha^{\rm ret}-F_\alpha^{\rm adv}\big),
\end{equation}
where $F_\alpha^{\rm ret}\equiv F_\alpha$ is the SF discussed so far, i.e., the one attributed to back-reaction from the retarded scalar field, and $F_\alpha^{\rm adv}$ is an ``advanced'' SF, constructed in just the same way from the {\em advanced} scalar field, i.e, the (unique) solution to the field equation satisfying advanced boundary conditions. We have $F_\alpha=F_\alpha^{\rm cons}+F_\alpha^{\rm diss}$. Then $\delta\chi_{\rm cons/diss}$ are defined by replacing $F_\alpha\to F_\alpha^{\rm cons/diss}$ in Eq.~(\ref{deltaphi1_final_method2}), respectively. 

Thanks to the time symmetry of geodesics in Schwarzschild spacetime, it is not necessary to compute $F_\alpha^{\rm adv}$ in practice. It can be shown (see, e.g., \cite{Barack:2009ux}) that $F_\alpha^{\rm cons}$ and $F_\alpha^{\rm diss}$ can be more readily extracted by combining information from the inbound and outbound legs of the orbit:
\begin{equation}\label{F_cons_diss}
F_\alpha^{\rm cons}(\tau) = \frac{1}{2}\Big(F_\alpha(\tau)-F_\alpha(-\tau)\Big), \qquad \qquad
F_\alpha^{\rm diss}(\tau) = \frac{1}{2}\Big(F_\alpha(\tau)+F_\alpha(-\tau)\Big),
\end{equation}
for the $\alpha = \{t,\varphi\}$ components. Explicit, simplified integral expressions for $\delta\chi^{\rm cons}$ and $\delta\chi^{\rm diss}$ are given in Eqs.\ (112) and (115) of Ref.~\cite{Barack:2022pde}.

It should be noted that our $\delta\chi$ is computed in a frame attached to $m_2$. In our scalar-charge model, however, this coincides with the center-of mass frame of the binary system, since we are ignoring the gravitational effect of the test mass $m_1$. 

In Sec.\ \ref{sec:NumericalSF} we shall review our numerical implementation of Eq.\ (\ref{deltaphi1_final_method2}) to compute $\delta\chi$ given $v$ and $b$. 

\section{Amplitudes-based methods for PM expansion}
\label{sec:PM}

\subsection{General setup}
\label{subsec:amp_generalities}

The relative velocity of two massive classical bodies on hyperbolic orbits can be close to the speed of light, not being bounded by the virial theorem. Thus, in the weak-field regime, the relevant framework for generic classical scattering events is the post-Minkowskian expansion, in which physical quantities are evaluated as a series expansion in Newton's gravitational constant $G$ while keeping all orders in the bodies' velocities compatible with Lorentz symmetry. This mirrors the usual expansion of scattering amplitudes in relativistic quantum field theory.  By extracting the classical limit of quantum scattering amplitudes, the major advances in this field provide an efficient route to study the classical scattering problem to high orders in $G$ \cite{Neill:2013wsa, Cheung:2018wkq, Bern:2019nnu, Bern:2019crd}.

Scattering amplitudes are basic quantities in quantum field theory. Following a well-defined procedure for identifying and extracting their classical part, they can then be used to calculate classical observables.  Here we will not describe the procedure in detail but will refer the reader to the literature~\cite{Kosower:2018adc, Bern:2019crd}. Several amplitudes-based approaches have been proposed: matching of amplitudes to those of an effective field theory leading to the two-body classical Hamiltonian~\cite{Cheung:2018wkq, Bern:2019nnu, Bern:2019crd}, eikonal methods~\cite{DiVecchia:2021bdo}, the relation between amplitude and the radial action~\cite{Bern:2021dqo, Bern:2021yeh}, and the observable-based approach pioneered by Kosower, Maybee, and O'Connell (KMOC)~\cite{Kosower:2018adc}. Heavy-particle effective theories~\cite{Damgaard:2021ipf, Brandhuber:2021eyq} provide possible shortcuts to the derivation of the requisite scattering amplitudes.
In addition, there are also worldline approaches~\cite{Kalin:2020mvi, Mogull:2020sak}, extending to the post-Mikowskian regime the NRGR framework of Goldberger and Rothstein~\cite{Goldberger:2004jt}. 
Classical quantities extracted from hyperbolic scattering can, in certain cases, be straightforwardly analytically continued to the phenomenologically-relevant case of bound-orbit dynamics~\cite{ Bern:2019crd, Cristofoli:2019neg, Kalin:2019rwq}.\footnote{At $\mathcal \calO(G^4)$ in  General Relativity, the tail effect makes it nontrivial to analytically continue between unbound and bound cases~\cite{Bini:2017wfr, Cho:2021arx}.} 

A quantum mechanical approach to observables makes use of notions, such as unitarity, that are not directly used in classical calculations. Harnessing the constraints they impose may render such calculations easier, as noted long ago by Kovacs and Thorne~\cite{Kovacs:1978IV}: ``Any classical problem can be solved quantum-mechanically; and sometimes the quantum solution is easier than the classical.''  The same authors commented on possible applications of Feynman diagrammatic methods to problems concerning gravitational radiation and that for classical macroscopic objects the naive dimensionless perturbative expansion parameter,
\begin{align}
   g = \frac{G m_1 m_2}{\hbar c} \approx \frac{m_1 m_2}{(10^{-8}\, {\rm kG})^2} ,
\end{align}
is much larger than unity. The classical limit has, however, a different effective coupling.
Bohr's correspondence principle states that classical physics emerges from the quantum theory in the limit of macroscopic conserved charges such as masses, electric charges, spins, orbital angular momenta, etc. In this limit the effective expansion parameter depends on the angular momentum $J$ of the two-body system,
\begin{align}
   g_\text{eff} = \frac{G m_1 m_2}{J c}  ,
\end{align}
which on the one hand is independent of Planck's constant and on the other can be small. Without further specifications on the magnitude of the velocity relative to the speed of light, this sets the expansion in the post-Minkowskian (relativistic weak field) regime. That is, two classical objects of typical Schwarzschild radii,
\begin{align}
 R_s = 2 G m,
\end{align}
are widely separated in impact parameter space $b$ at a fixed velocity, i.e. $R_s \ll b$.\footnote{More generally, the minimal distance between the two bodies is governed by $ b v^2 \sim J v/m$, which should be much larger than $R_s$ in the weak field regime.} This in turn implies that, in the classical limit, the Fourier-conjugate of the separation of the two particles---the momentum transfer---is much smaller than the characteristic incoming and outgoing momenta, corresponding to a soft expansion.

With the goal of identifying the classical part of quantum scattering amplitudes (and constructing only that part), let us briefly review the dependence of four-point amplitudes on the impact parameter and on the Compton wavelength $\lambda_c \sim \hbar$. 
As has been explained in e.g.\ Ref.~\cite{Kosower:2018adc}, amplitudes are ever more singular in the classical limit $\hbar \to 0$ at higher orders in perturbation theory.  This may be understood intuitively by recalling that in nonrelativistic quantum mechanics, the $n$-th order correction to scattering amplitudes contains contributions from $n$ insertions of the leading-order interaction potential. Thus, if the tree-level scattering amplitude is classical ${\cal O}(\hbar^{-1})$, then the complete $L$-loop classical amplitude 
contains classically-singular terms up to ${\cal O}(\hbar^{-(L+1)})$, as dictated by the semiclassical approximation:
\begin{align}
i {\cal M} \sim 
e^{i S/\hbar } -1 
\sim \frac{1}{\hbar} \sum_L \left(\frac{G m_1 m_2}{J}\right)^{L+1} \sum_{l\ge -L} {\cal C}_{l} \left(\frac{\lambda_c}{b}\right)^l ,
\label{classical0}
\end{align}
where ${\cal C}_{l}$ are some coefficients;  recall that the Compton wavelength is $\lambda_c\sim{\cal O}(\hbar)$. 

As in nonrelativistic quantum mechanics, all these classically-singular terms, ${\cal C}_{l}$ with $l<0$, are determined by lower-loop amplitudes and are subtracted out by various means in order to identify the classical part of scattering amplitudes. Up to terms depending on the details of the subtraction procedure, this classical amplitude is determined by ${\cal C}_{0}$.  The terms containing positive powers of the Compton wavelength are quantum mechanical, and therefore not immediately relevant to classical physics.

\begin{figure}[tbh]
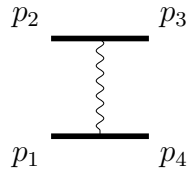

\centering {\treet}
\caption{ The lowest order Feynman diagram describing the scattering of two point particles.}
        \label{fig:ScalarFeynman}
\end{figure}

From a practical computational point of view, it is convenient to realize the expansion \eqref{classical0} in the momentum space form of the $2\to2$ scattering process of two point-like heavy objects exchanging massless mediators, as shown at lowest order in \fig{fig:ScalarFeynman}.
We use an all-outgoing convention for external momenta in amplitudes in this paper.
As discussed above, the momentum transfer $q=p_4+p_1=-p_2-p_3$, which is Fourier-conjugate of the impact parameter, is much smaller than the ${\cal O}(1)$ external momenta, so taken to be $\calO(\hbar)$. 
Beyond tree level, we need to understand the scaling of mediator momenta in loop integrals. We do not expect macroscopic bodies to fluctuate far off shell, so every massless mediator must also have momentum with the same $\hbar$ scaling as the momentum transfer, i.e. ${\cal O}(\hbar)$. With this scaling one can show that closed loops of mediators 
can generate only quantum terms; this is in agreement with the intuitive picture that loops of mediators are quantum mechanical.
Furthermore, the classical part of the amplitude $\mathcal M_{\rm class.}$, related to $l = 0$ in Eq.~\eqref{classical0}, depends on the momentum transfer $q$ as
\begin{align}
\sim \frac{G^{L+1}}{|q|^{2-L}} \, (\ln |q|)^x,  \hskip 1 cm 
x = \left\{
\begin{array}{ll}
0 & L\text{ odd,} \cr
1 & L\text{ even}, \; L >0 .
\end{array}
\right.
\label{classical}
\end{align}
The procedure described here is implemented by the so-called soft expansion in the method of regions pioneered by Beneke and Smirnov~\cite{Beneke:1997zp}. It is sometimes advantageous to split the classical region further into instantaneous exchanges of `potential modes' that are relevant to conservative dynamics, and on-shell `radiation modes' with relevance to dissipative effects. For more details on the kinematic definitions of the regions, see e.g.~Ref.~\cite{Beneke:1997zp, \citeNRGR, Smirnov:2004ym, Bern:2019crd, Bern:2021yeh}.

Eqs.~\eqref{classical0} and \eqref{classical} emphasize that determination of higher PM orders of classical amplitudes requires the evaluation of multi-loop scattering amplitudes in the classical limit and therefore profits from the extensive developments in this field, including the construction of amplitude integrands using unitarity-based methods~\cite{Bern:1994zx, Bern:1994cg, Bern:1995db, Bern:1997sc, Britto:2004nc}, efficient integral reduction methods using integration-by-parts relations~\cite{Chetyrkin:1981qh, Laporta:2000dsw} implemented in automated programs (we use Refs.~\cite{Smirnov:2008iw, Smirnov:2019qkx}), and the evaluation of the resulting `master' integrals via differential equations methods~\cite{Kotikov:1990kg, Bern:1993kr, Remiddi:1997ny, Gehrmann:1999as}. 

In subsequent sections, we leverage these advances and target the classical scattering angle in the PM expansion in a quantum field theory model of massive scalars, with forces mediated by a massless scalar and gravity. 
Its Lagrangian is given in Eq.~\eqref{eq:model_action_minimal} below, and it mirrors the model used in the SF calculations described in the previous section. We present results for the scattering angle through order $G^3 q_s$ and compare them with the results of the self-force calculations. In our discussion we consider separately the conservative and radiative dynamics; to this end, it is useful to separate the soft mediator modes into potential modes, whose four-momenta scale as $k\sim \left(q v, q \right)$,
and radiation modes whose momenta scale as  
$k\sim \left( q v, q v \right)$,
%
where the two entries in each represent the energy and three-momentum scalings. Conservative dynamics contains the effect of potential modes, which are off shell and thus cannot be radiated, and the tail effect from radiation modes which are emitted and reabsorbed by the system and do not escape to infinity.  
Dissipative dynamics is due only to radiation. In the language of the method of regions \cite{Beneke:1997zp} for loop integrals, oddness under time reversal is a diagnostic of dissipative effects due to the presence of radiation modes. 
Restricting to 1SF but all orders in PM~\cite{Barack:2022pde}, the conservative and dissipative parts of the scattering angle directly correspond to the  terms that are even and odd in velocity.\footnote{
It can be shown that the conservative force at 1SF has a Hamiltonian description~\cite{Isoyama:2014mja,Fujita:2016igj,Blanco:2022mgd}.
}
Beyond 1SF order, however, there is no unique separation between the conservative and dissipative parts; only the complete results are well-defined.
See Eq.~(12.32) of Ref.~\cite{Bini:2022enm}
and the complete calculation at 4PM in Ref.~\cite{Dlapa:2022lmu,Dlapa:2023hsl}.
But this subtlety is beyond the scope of the present paper.

We follow the amplitudes-based approach of Refs.~\cite{Bern:2021dqo, Bern:2021yeh} to conservative dynamics and construct the radial action.  As in Ref.~\cite{Bern:2021yeh}, we identify the radiation-mode contribution to the conservative dynamics as the real part of the corresponding amplitude with the Feynman-$i\varepsilon$ prescription for the radiation modes of the messengers. This is equivalent to the prescription of Ref.~\cite{Bini:2021gat} using principal-value propagators, which in turn corresponds to time-symmetric propagators~\cite{Wheeler:1949hn, Damour:1995kt, Damour:2016gwp}.
We find the dissipative contributions to the scattering angle, for the model in Eq.~\eqref{eq:model_action_minimal} below, using the linear response approach of Ref.~\cite{Bini:2012ji,Damour:2020tta} which takes as input the  total energy and angular momentum loss at 3PM order.
We will discuss aspects of these methods in Secs.~\ref{subsec:PM_cons} and~\ref{subsec:PM_rad}, respectively.


\subsection{EFT description and tidal effects}
\label{subsec:EFT_and_tidal}

Underlying the scattering amplitudes approach is an effective field theory description of two black holes interacting via gravity. In this approach, the two black holes, which we assume to be spinless, are described by point particles created by massive scalar fields $\phi_1$ and $\phi_2$.  This mimics the standard approach to dynamics in both the PN and PM frameworks, where a separation of scales is assumed allowing us to treat the black holes as point particles. Any internal structure is described by a set of tidal 
operators whose coefficients can be fixed by matching EFT amplitudes with amplitudes
computed e.g. using black hole perturbation theory. 
%
Furthermore, to match results from the self-force approach we take $m_1\ll m_2$ and expand in the ratio of masses. Furthermore, we take the scalar fields corresponding to the lighter black hole, $\phi_1$, to carry a charge, $Q$, which interacts with a massless long-range scalar field, $\psi$.
%
\begin{figure}[tb!]
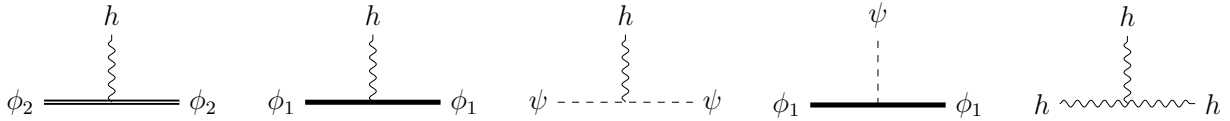

\centering
    $
    \vcenter{\hbox{\scalebox{.9}{\threeptPPh}}} \hskip .3 cm
    \vcenter{\hbox{\scalebox{.9}{\threeptpph}}} \hskip .3 cm
    \vcenter{\hbox{\scalebox{.9}{\threeptcch}}} \hskip .3 cm
    \vcenter{\hbox{\scalebox{.9}{\threeptppc}}} \hskip .3 cm
    \vcenter{\hbox{\scalebox{.9}{\threepthhh}}}
    $
\caption{The three-point interaction vertices of the model defined by the action in Eq.~(\ref{eq:model_action_minimal}). The  double line represents the heavy black hole field $\phi_2$, the single solid line denotes the light charged scalar field $\phi_1$. The wavy line represents gravitons, and the dashed line denotes the self-force scalar $\psi$.}
\label{fig:ScalarVertex}
\end{figure}

Our minimal model ignoring any tidal effects is defined by the effective field theory action
\begin{align}
\label{eq:model_action_minimal}
S \!= \!\int d^D x \sqrt{-{\sf g}} \biggl[\tfrac{R}{16\pi G}
+\tfrac{1}{2}  \phi_1 (\Box - m_1^2)  \phi_1 
+\tfrac{1}{2}  \phi_2 (\Box - m_2^2) \phi_2  
+\tfrac{1}{2}   \psi \,\Box \, \psi  
-2 \sqrt{\pi} \, m_1 Q  \, \psi \phi_1^2 
\biggr] .
\end{align}
Here ${\sf g}_{\mu\nu}$ is a weak-field metric expanded in perturbations around flat space ${\sf g}_{\mu \nu} = \eta_{\mu \nu} + \sqrt{32\pi G}\, h_{\mu\nu}$, ${\sf g}$ is its determinant and $R$ is the corresponding Ricci-scalar. In dimensional regularization, $D=4-2\epsilon$ is the space-time dimension, and the normalization of the scalar charge $Q$ has been chosen to match the conventions in Refs.~\cite{Gralla:2021qaf, Barack:2022pde}. Here the massless Klein-Gordon scalar $\psi$ couples minimally to gravity and also couples directly to the light black hole field $\phi_1$ but only gravitationally to the heavy black hole field $\phi_2$. The three-point Feynman vertices following from the Lagrangian \eqref{eq:model_action_minimal} are shown in Fig.~\ref{fig:ScalarVertex}.

The massless scalar field $\psi$ is related to $\Psi$ of \Sec{sec:ScalarSF} by a generically nontrivial field redefinition. Nonetheless, scattering amplitudes and other classical observables are field-redefinition invariant, so we will not need to be concerned about the precise mapping between $\Psi$ and $\psi$.

The EFT above is only valid at distances much larger than the Schwarzschild radii of the black holes, $r\gg R_{s, i}= 2G m_i$, also known as the \emph{far zone}, where the point-particle approximation is valid. Equivalently, the EFT describes the scattering of long-wavelength, i.e. $G m_i\omega\ll1$, waves off the black holes. The physics in the \emph{near zone}, $r\sim 1/\omega \lesssim G m_i$, includes the tidal properties of the black holes. In particular, the black holes' response to the massless scalar $\psi$ is encoded in nonminimal couplings beyond the terms already present in \Eq{eq:model_action_minimal}, which take the schematic form\footnote{All independent operators (i.e. operators with distinct matrix elements) describing the relevant physics should be present in the EFT action. Since we are interested in the 1SF order, operators with $\phi_1$ and $\psi$ cannot be present, while on dimensional grounds operators with either of the massive scalars and two curvature tensors are of too high a dimension to contribute to contribute to the orders we will be evaluating~\cite{Goldberger:2004jt}.    
We are thus left with operators involving $\phi_2$ and $\psi$. In the absence of $\phi_1$, constant shifts of the scalar $\psi$ in the action \eqref{eq:model_action_minimal} are a symmetry; the nonminimal couplings of $\phi_2$ to $\psi$ are expected to respect this symmetry and therefore should involve only derivatives of $\psi$, thereby ruling out operators such as $\phi^2_2\, \psi^2$.} 
\begin{align}
\label{eq:tidal_operators0}
 \mathcal{O}^{{\rm tidal}} \sim c_{ab} (\partial^a \phi_2)^2 (\partial^b\psi)^2 ,
\end{align}
where the $c_{ab}$ are Wilson coefficients (closely related to Love numbers) labeled by $a$ and $b$. The power counting of such tidal effects can be determined by considering classical gravitational scattering of $\psi$ off a massive black hole (represented by $\phi_2$) in full General Relativity. The corresponding amplitude can be computed by solving the Regge-Wheeler equation with appropriate boundary conditions at $r\to\infty$ and the horizon; it schematically reads,
\begin{align}
\calM_{\rm GR} \sim  G m_2^2 [1 + G m_2 \omega +  (G m_2 \omega)^2 + \cdots ].
\end{align}
The minimal EFT in \eqref{eq:model_action_minimal} cannot reproduce the full expansion in powers of $G m_2 \omega$, but the mismatch can be accounted for by the contribution of the tidal operators in \Eq{eq:tidal_operators0} which lead to amplitudes of the form
\begin{align}
\calM^{{\rm tidal}}_{{\rm EFT}} \sim c_{ab}\, m^{2a}_2 \, \omega^{2b}.
\end{align}
This leads to the power-counting
\begin{align}
 c_{ab} \sim G^{1+2b} \, m^{2(1-a+b)}_2 .
\end{align}
Hence the leading tidal effects are captured by adding to the effective action
\begin{align}
S^{\rm tidal} = &
  G^3 \! \int\! d^D x \sqrt{-{\sf g}}
    \Big[ 
   (4\pi c_1)\left[m^2_2(\partial_\mu\phi_2 \partial^\mu\psi)^2 {-} m^4_2 \phi_2^2 (\partial_\mu \psi)(\partial^\mu\psi)\right] \nonumber
   \\
&\qquad\qquad\qquad~
+(4\pi c^{{\rm bare}}_2)\, m_2^2\, (\partial_\mu\phi_2 \partial^\mu\psi)^2
   \Big] 
   + \calO(G^4),
\label{eq:tidal_operators}
\end{align}
where  $c_1$ and $c^{{\rm bare}}_2$ are dimensionless Wilson coefficients \cite{Cheung:2020sdj, Bern:2020uwk} and $\calO(G^4)$ denotes additional higher dimensional tidal operators with more derivatives on $\psi$ and, at higher orders $G$, also for gravitational tidal operators. A novel feature of the scalar model is that the finite-size effect appears at a lower order than in standard GR, where the leading tidal operators scale schematically as $G^5 \phi_2^2 R^2$ \cite{Goldberger:2004jt} and the gravitons  take the place of the derivatively-coupled scalar field $\psi$ here. Following standard EFT procedures, the dimensionless Wilson coefficients need to be determined by a matching calculation akin to the one performed by e.g.~Refs.~\cite{Kol:2011vg, Chia:2020yla, Hui:2020xxx, Charalambous:2021mea, Hui:2021vcv, Ivanov:2022qqt, Ivanov:2022hlo}.  As we shall see, the second operator in Eq.~\eqref{eq:tidal_operators}, with coefficient $c^{{\rm bare}}_2$, is necessary for the consistency of the EFT, because it is required to absorb an ultraviolet divergence, thereby also explaining the superscript `bare' introduced above. The first operator, with coefficient $c_1$, does not cancel a divergence of the minimal amplitude but it is allowed by the symmetries of the theory and therefore its coefficient should be determined through a matching computation. 
We have written the operators in  Eq.~\eqref{eq:tidal_operators} in a 
basis that explicitly decomposes them into the analogs of static and dynamic 
QFT Love numbers, respectively.  
Refs.~\cite{Kol:2011vg, Chia:2020yla, Hui:2020xxx, Charalambous:2021mea, Hui:2021vcv, Ivanov:2022qqt, Ivanov:2022hlo} carried out such a matching in a worldline EFT version of the scalar model and found that the static Love number, which should be equivalent to our $c_1$, vanishes. This result has also been confirmed through the GR calculation of the deformation of a spherical body by an external tidal field \cite{Binnington:2009bb}. To confirm that our $c_1$ indeed vanishes we would need to carry out a similar matching in the context of our formalism, as it can in principle differ from others by scheme choices and field redefinitions that might lead to finite shifts of Wilson coefficients. Such a computation requires a nontrivial extension of Refs.~\cite{Kol:2011vg, Chia:2020yla, Hui:2020xxx, Charalambous:2021mea, Hui:2021vcv, Ivanov:2022qqt, Ivanov:2022hlo} beyond the static sector and is left as interesting future work.   
In this work, instead of performing a proper matching calculation, we will estimate the coefficients by comparing them to the numerical self-force computation. Below we shall see that within the relatively large uncertainties that occur as we push the comparison between SF and PM to its limits, the result is compatible with the expectation that the scalar model static Love number vanishes.

\subsection{Conservative Dynamics}
\label{subsec:PM_cons}

\begin{figure}[tb]
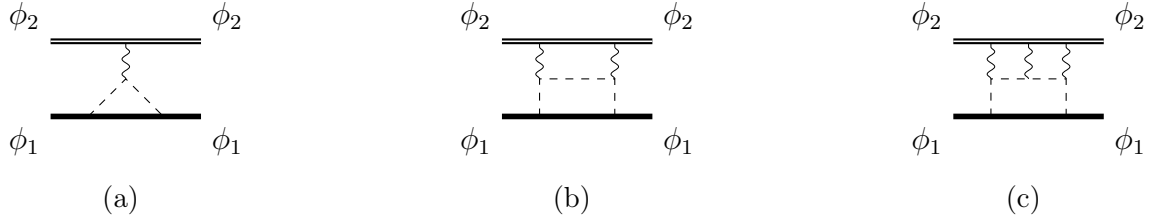

    \centering
      \begin{subfigure}{0.3\textwidth}\centering
       $\vcenter{\hbox{\scalebox{1}{\twoPMeg}}}$
        \caption{ }
        \label{fig:2pm_eg_diag}
      \end{subfigure}
      \hfill
      \begin{subfigure}{0.3\textwidth}\centering
      $\vcenter{\hbox{\scalebox{1}{\threePMeg}}}$
      \caption{}
      \label{fig:3pm_eg_diag}
      \end{subfigure}
      \hfill
      \begin{subfigure}{0.3\textwidth}\centering
      $\vcenter{\hbox{\scalebox{1}{\fourPMeg}}}$
      \caption{}
      \label{fig:4pm_eg_diag}
      \end{subfigure}
\caption{
\label{fig:OneLoopDiag_scalar}
Representative diagrams at $\calO(Q^2)$ and at (a) 2PM, (b) 3PM, and (c) 4PM orders.}
\end{figure}

Following standard procedures, the Lagrangian generates a set of Feynman rules, with the three-point vertices of the model shown in \fig{fig:ScalarVertex}.  The resulting expressions for the terms proportional to $q_s$, see sample diagrams in Fig.~\ref{fig:OneLoopDiag_scalar}, are much simpler than for the corresponding purely gravitational case and can be straightforwardly evaluated via the Feynman rules to give an integrand for the terms in the scattering amplitude proportional to $q_s$.   Following the simple scaling rules discussed in Sec.~\ref{subsec:amp_generalities}, the quantum terms are removed leaving an integrand which is then integrated using the standard tools of integration by parts~\cite{Chetyrkin:1981qh, Laporta:2000dsw} and differential equations~\cite{Kotikov:1990kg, Bern:1993kr, Remiddi:1997ny, Gehrmann:1999as}. This process is enormously simplified in the classical limit with the important observation that a good set of variables effectively leaves only single-scale integrals to all orders of perturbation theory~\cite{Parra-Martinez:2020dzs}.  The resulting amplitude can then be matched to a two-body effective field theory that extracts the potential and Hamiltonian~\cite{Cheung:2018wkq} from which observables can be found by solving Hamilton's equations.

Here we follow a more direct route to the scattering angle using an identification between the scattering amplitudes and the radial action~\cite{Bern:2021dqo}. The radial action is a basic object in classical mechanics. In impact-parameter space it is given by the integral of the radial momentum $p_r$ along the scattering trajectory (with appropriate regularization of the long-distance contribution), 
\begin{equation}
I_r(J) = \int_\text{traj.} p_r \, dr\, .
\end{equation}
The conservative scattering angle is obtained via
\begin{align}
    \chi^{\rm cons} = - \frac{\partial I_r(J)}{\partial J}\, ;
\end{align}
similar thermodynamic-like relations exist for other observables. To state the relationship between the scattering amplitude and the radial action we define the momentum-space radial action $I_r(q)$ as Fourier transform of the radial action in impact parameter space, 
\begin{equation}
I_r ( q) 
= 4m_1 m_2 \sqrt{\sigma^2-1} \int  {d^{D-2}\bm b} \, \mu^{-2\epsilon} \, e^{i\bm q\cdot \bm b}\, I_r(J) .
\label{eq:FT_J}
\end{equation}
As before, $\epsilon$ is the dimensional-regularization parameter, $\bm q$ is the space-like part of the transferred momentum $q$, $b\equiv |\bm b|= {J}/{(m_1 \sqrt{\sigma^2-1})}$ is the impact parameter in the COM frame, and $\mu$ is the usual scale that appears in dimensional regularization. 

With this definition, the amplitude through classical order and the momentum-space radial action are related by~\cite{Bern:2021dqo} 
\begin{align}
\label{eq:amp_rad_act}
	{\cal M}_1( q) &= I_{r,1}( q),  \nonumber \\ 
	{\cal M}_2( q) &= I_{r,2}( q) 
        + \int_{\bm \ell}\frac{n^{(1)}_2}{Z_1} , \nonumber  \\
	{\cal M}_3( q) &= I_{r,3}( q) 
    + \int_{\bm \ell}\frac{n^{(2)}_3}{Z_1 Z_2} 
    + \int_{\bm \ell}\frac{n^{(1)}_3}{Z_1} ,
 \\
 {\cal M}_4( q) &= 
    I_{r, 4}( q)    
    + \int_{\bm \ell} \frac{n^{(3)}_4}{Z_1  Z_2  Z_3 }
    + \int_{\bm \ell} \frac{n^{(2)}_4}{Z_1  Z_2 }
    + \int_{\bm \ell} \frac{n^{(1)}_4}{Z_1} ,
    \nonumber
\end{align} 
where ${\cal M}_n(q)$ is the ${\cal O}(G^n)$ semi-classical amplitude, 
and $I_{r, n}( q)$ is the momentum-space radial action at the same order.
The $\bm\ell$ integration in Eq.~\eqref{eq:amp_rad_act} is defined as
\begin{equation}
\int_{\bm \ell} \equiv \int \prod_{i=1}^n \,\frac{d^{D-1}\bm \ell_i}{(2\pi)^{D-1}}\, (2\pi)^{D-1} \delta(\sum_{j=1}^n\bm\ell_j-\bm q) \ .
\end{equation}
${\cal M}_n(q)$ is computed following Ref.~\cite{Bern:2021dqo}, by expanding the matter poles about the momentum component in the direction of the spatial component of $\bar{p}_{1} :=(p_4-p_1)/2$, which we take to be along the $\hat{\bm z}$ unit vector, resulting in a direct relationship between the real part of the classical scattering amplitude and the radial action. The terms containing the denominator factors 
\begin{align}
&Z_j := - 4 m_1 m_2 \sqrt{\sigma^2-1} \big((\bm \ell_{1}+\bm \ell_2+\dots +\bm\ell_{j})\cdot \hat{\bm z}+i\varepsilon \big) \,,
\end{align}
tag the iteration terms, which are uninteresting because they contain only lower-order information. The Feynman $i\varepsilon$ prescription is included to specify the boundary condition and is inherited from that of the matter propagators prior to the expansion. 
Further details on these iteration terms and the structure of the numerators $n_i^{(j)}$ are given in Ref.~\cite{Bern:2021dqo}. This organization of the amplitude is inspired by the well-known eikonal expansion~\cite{Glauber:1987bb, Amati:1990xe, Saotome:2012vy, Akhoury:2013yua, DiVecchia:2019myk, DiVecchia:2019kta, Bern:2020gjj, DiVecchia:2020ymx, DiVecchia:2021bdo}, with an important difference that this prescription manifests the pole structure in $Z_j$ of the amplitude's integrand and therefore each term has definite scaling in the soft expansion so the classical part can be isolated.  The net effect is that iteration terms can be dropped without explicitly evaluating them~\cite{Bern:2021dqo}. 

The direct relationship between the classical part of the amplitude and the radial action relies on the specific treatment of the poles and differs from the earlier treatment~\cite{Cheung:2018wkq, Bern:2019nnu, Bern:2019crd}. Ref.~\cite{Bern:2021dqo} explicitly verified \Eq{eq:amp_rad_act} through $\mathcal{O}(G^4)$ by comparing the amplitude calculation in EFT to the radial action from classical mechanics. Implicitly, we keep only the real part of the scattering amplitudes.

\subsection{Radiative Dynamics}
\label{subsec:PM_rad}

We use the Bini-Damour linear response formula \cite{Bini:2012ji, Damour:2020tta} 
\begin{align}
	\chi^{\rm diss} = 
 \frac{\partial \chi^{\rm cons}}{\partial J}\left(-\frac{1}{2} J_{\rm rad}\right)
	+\frac{\partial  \chi^{\rm cons}}{\partial E}\left(-\frac{1}{2} E_{\rm rad}\right),
\end{align}
which is valid to linear order in dissipation.
Here the  radiated angular momentum $J_{\rm rad}$ and energy $E_{\rm rad}$  are  expanded into a scalar-field self-force contribution and others from gravitational self-force
\begin{align}
 J_{\mathrm{rad}} =  {}& q_{s}\,\delta J_{\mathrm{rad}} +\mathcal{O}(q_s^2, q_m) ,\\
  E_{\mathrm{rad}} =  {}& q_{s}\,\delta E_{\mathrm{rad}} +\mathcal{O}(q_s^2, q_m) .
\end{align}
Since we are only interested in the first-order SF, we only need $\delta J_{\rm rad}$ and $\delta E_{\rm rad}$ emitted by the charge on a geodesic scattering orbit. This linear order in dissipation captures the odd-in-velocity sector of the scattering angle. Note that this linear response formula gives the scattering angle in the initial center-of-mass frame. 
However, since the back-reaction on the heavy black hole is suppressed by a factor of the small mass ratio, the effect on the scattering angle is of the same order as that of the gravitational self-force, which we anyway neglect in our analysis. Thus, in our approximation, the scattering angle in the center-of-mass frame is the same as it is in the rest frame of the heavy object. 
See Appendix~\ref{sec:Frames} for a more complete discussion of this point. 

To calculate the angular momentum and energy carried away in the scalar-field waves, we start with the scalar waveform in the asymptotic region,
evaluated at $x^\mu=(t,r \bm {\hat{r}})$:
\begin{align}
	\psi (x) 
	&= \int \widetilde{dk} \left(i J(k) e^{-ik\cdot x} 
    -iJ^*(k) e^{ik\cdot x}  \right)
  \nonumber\\
    &= \frac{1}{4\pi r}\int \frac{d\omega}{2\pi} \left( J(k) e^{-i\omega (t-r)} +\textrm{c.c.} \right)\big\vert_{k=\omega(1,\hat{\bm r})}
    +\mathcal{O}\left(r^{-2}\right),
    \label{eq:phi_mode} 
\end{align}
Here $\widetilde{dk} = {d^3k}/{((2\pi)^3 2\omega )}$ is the Lorentz-invariant on-shell phase space measure and $J(k)$ is the classical scalar current in momentum space.\footnote{The field equation in position space reads $\nabla^\alpha \nabla_\alpha \psi(x) = -J(x)$.}
The first line is the scalar field written in on-shell momentum space, and in the second line, we expand in the asymptotic region at a large distance $r$ but with finite $t-r$ and momentum $k^\mu$ is evaluated at $k^\mu=\omega(1,\hat{\bm r})$.
Given the stress-energy tensor of the scalar $T^{\mu\nu}=\partial^\mu \psi \partial^\nu \psi -\frac{1}{2}g^{\mu\nu} (\partial \psi)^2$, we obtain the radiated linear and angular momentum following the procedure in Ref.~\cite{Manohar:2022dea}:
\begin{align}
	P^\mu_{\rm rad}
	&= \int \widetilde{d k} k^\mu J^*(k) J(k), \\
    J^{\mu\nu}_{\rm rad} &= \int \widetilde{dk}\, J(k)^* \left(i k^{\mu} \frac{\partial}{\partial k_{\nu}}-
    i k^{\nu} \frac{\partial}{\partial k_{\mu}}\right) J(k).
    \label{eq:EJrad_Genformula}
\end{align}
In our frame choice, $E_{\rm rad} = P^0_{\rm rad}$ and $J_{\rm rad} \equiv J^{12}_{\rm rad}$. Following Refs.~\cite{Herrmann:2021lqe, Herrmann:2021tct} we recast the phase-space integrals into multi-loop integrals by reverse unitarity~\cite{Anastasiou:2002yz}. Both the radiated energy~\cite{Herrmann:2021lqe, Herrmann:2021tct} and angular momentum~\cite{Manohar:2022dea} have been calculated to 3PM order in GR.

\begin{figure}[tb]
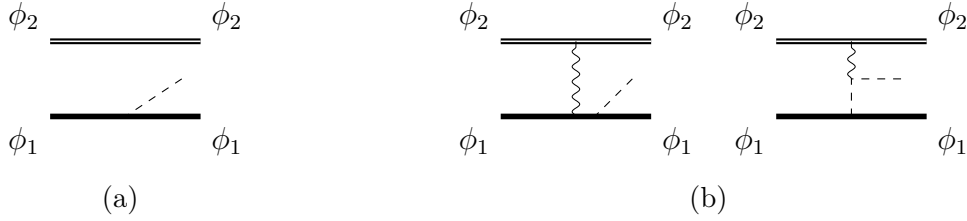

    \centering
      \begin{subfigure}{0.3\textwidth}\centering
       $\vcenter{\hbox{\scalebox{1}{\zeroPMradQ}}}$
        \caption{}
      \end{subfigure}
      \qquad
      \begin{subfigure}{0.5\textwidth}\centering
      $\vcenter{\hbox{\scalebox{1}{\onePMradQEgOne}}}$
      $\vcenter{\hbox{\scalebox{1}{\onePMradQEgTwo}}}$
      \caption{}
      \end{subfigure}
      \hfill
\caption{
\label{fig:5pt_scalar}
Representative diagrams at $\calO(Q)$ and (a) 0PM and (b) 1PM order}
\end{figure}

While the energy can only be emitted by $\omega\ne 0$ scalar waves, the radiated angular momentum receives contributions from both the finite frequency and the zero-frequency limit. In this limit, we can approximate the particle trajectory as~\footnote{
The kink at $\tau=0$ can be chosen as the time at the
periastron, but the result is independent from this choice since we integrate the full trajectory from $\tau=-\infty$ to $\tau=\infty$.
}
\begin{align}
    x^\mu(\tau) &= u^\mu_i\tau + \theta(\tau)(u^\mu_f - u^\mu_i)\tau,
\end{align}
where $u^\mu_i$ and $u^\mu_f$ are the initial and final velocity of the particle.
In an all-outgoing convention, $m_1 u_i^\mu= -p_1^\mu=(E_1,\bm p_1)$ and $m_1 u_f^\mu = p_4^\mu=(E_4,\bm p_4)$ are the 4-momenta of $\phi_1$.
This leads to  
\begin{align}
	J(k)|_{\omega \rightarrow 0^+}
	&= -2i\sqrt{\pi}Q m_1 \left[
    \frac{\hat{\delta}(\omega)}{2(E_1-\bm p_1\cdot \hat{\bm k})}
	+ \frac{i}{\omega+i\varepsilon} \,\left(
    \frac{1}{E_4-\bm p_4\cdot \hat{\bm k}}
    -\frac{1}{E_1-\bm p_1\cdot \hat{\bm k}}
    \right)
    \right],
    \label{eq:softlimit}
\end{align}
where we insert a factor of $1/2$ in the $\hat{\delta}(\omega)$ to account for the splitting of delta function into the positive frequency domain.
To absorb factors of $2\pi$, we define 
\begin{align}
 \hat{d} x = \frac{dx}{2\pi},
 \qquad 
 \hat{\delta}(x) = 2\pi \delta(x).
\end{align}
The interference between $\hat{\delta}(\omega)$ and $1/(\omega+i\varepsilon)$ leads to non-trivial radiated angular momentum in the zero-frequency limit.
The radiated angular momentum starting at 2PM is entirely due to this interference.
At higher orders, $\delta J_{\rm rad}$ receives contributions from both the zero-frequency limit and finite frequency waveform.
$\delta E_{\rm rad}$ also arises from the finite frequency waveform.
We obtain the finite-frequency waveform from the amplitude $\mathcal{M}(\phi_1,\phi_2,\phi_1,\phi_2,\psi)$ using the observable-based formalism~\cite{Kosower:2018adc}. To leading order, the relation between the amplitude (in all-outgoing momentum convention) and the classical scalar current reads
\begin{align}
    \hspace{-.5cm}
	J(k)
	{=} \int\! \hat{d}^D \ell_1 \hat{d}^D \ell_2 \hat{\delta}(2p_1\cdot \ell_1) \hat{\delta}(2p_2\cdot \ell_2)
	e^{i\ell_1\cdot b_1+i\ell_2\cdot b_2}
	\hat{\delta}^{(D)}(k+\ell_1+\ell_2) 
    \hspace{-1.3cm}
    \vcenter{\hbox{\scalebox{.9}{\FivePtTree}}}
    \hspace{-.6cm}
    \label{eq:waveformKMOC}
\end{align}
See \fig{fig:5pt_scalar} for sample diagrams that are needed to evaluate the dissipative contribution through the 4PM scattering angle.

Starting at 4PM, the dissipative angle has both odd-in-velocity and even-in-velocity contributions. 
The latter only arises when the radiation reaction force is even under time reversal or applied beyond the linear order.
As we reviewed in Sec.~\ref{subsec:amp_generalities},~none of these occurs at 1SF order. Therefore the linear response approach is sufficient to capture all the dissipative contributions needed for this paper.
The complete contribution beyond 1SF requires the full calculation in the observable-based formalism~\cite{Kosower:2018adc} or in an in-in-type approach~\cite{Dlapa:2022lmu,Jakobsen:2022psy, Dlapa:2023hsl}.

\section{Numerical self-force calculation}
\label{sec:NumericalSF}

Our numerical calculation of the SF correction to the scattering angle is performed using an adaptation of the code developed in Ref.\ \cite{Barack:2022pde}. The details of the numerical method are described in Sec.\ VIII and App.\ B of that work. Here we give a brief review of the method and then describe the adaptations made to enable the production of data suitable for our precision PM comparison. These were mostly incorporated as post-processing steps.

\subsection{Numerical integration of the scalar-field equations}

The main numerical task is the construction of a retarded solution to the sourced Klein-Gordon equation (\ref{eqn:Scalar3+1}), for a given sourcing scattering geodesic. We take advantage of the separability of the equation into multipole (spherical-harmonic) modes defined on spheres $r$=const around the Schwarzschild black hole, in order to reduce the equation to a set of hyperbolic evolution equations in 1+1 dimensions (time+radius), one for each multipolar mode. Specifically, we expand the field $\Psi$ in the form
\begin{equation}
\Psi = \frac{2\pi Q}{r} \sum_{\ell =0}^{\infty} \sum_{m=-\ell}^{\ell} \psi_{\ell m}(t,r) Y_{\ell m} (\theta, \varphi),
\end{equation}
where $Y_{\ell m}$ are standard spherical harmonics. Each of the modal time-radial functions $\psi_{\ell m}(t,r)$ then satisfies
\begin{align}\label{eqn:SourcedFieldEquation}
\frac{\partial^2\psi_{\ell m}}{\partial u \partial v}  + V(r) \psi_{\ell m}  =\frac{1}{2 \sigma r_p(t)} \left( 1-\frac{2m_2}{r_p(t)}\right)^2 \delta \left(r - r_p(t)\right) \bar{Y}_{\ell m}(\pi/2,\varphi_p(t)),
\end{align}
where $v \equiv t+r+2m_2\ln[r/(2m_2)-1]$ and $u \equiv t-r-2m_2\ln[r/(2m_2)-1]$ are the Eddington-Finkelstein advanced and retarded time coordinates, an overbar denotes complex conjugation, and
\begin{equation}
V(r) \equiv \frac{1}{4r^2} \left( 1-\frac{2m_2}{r}\right)\left(\ell(\ell+1) + \frac{2m_2}{r} \right).
\label{eq:ScalarPotential}
\end{equation}

\begin{figure}[h!]
\centering
\includegraphics[width=0.55\linewidth]{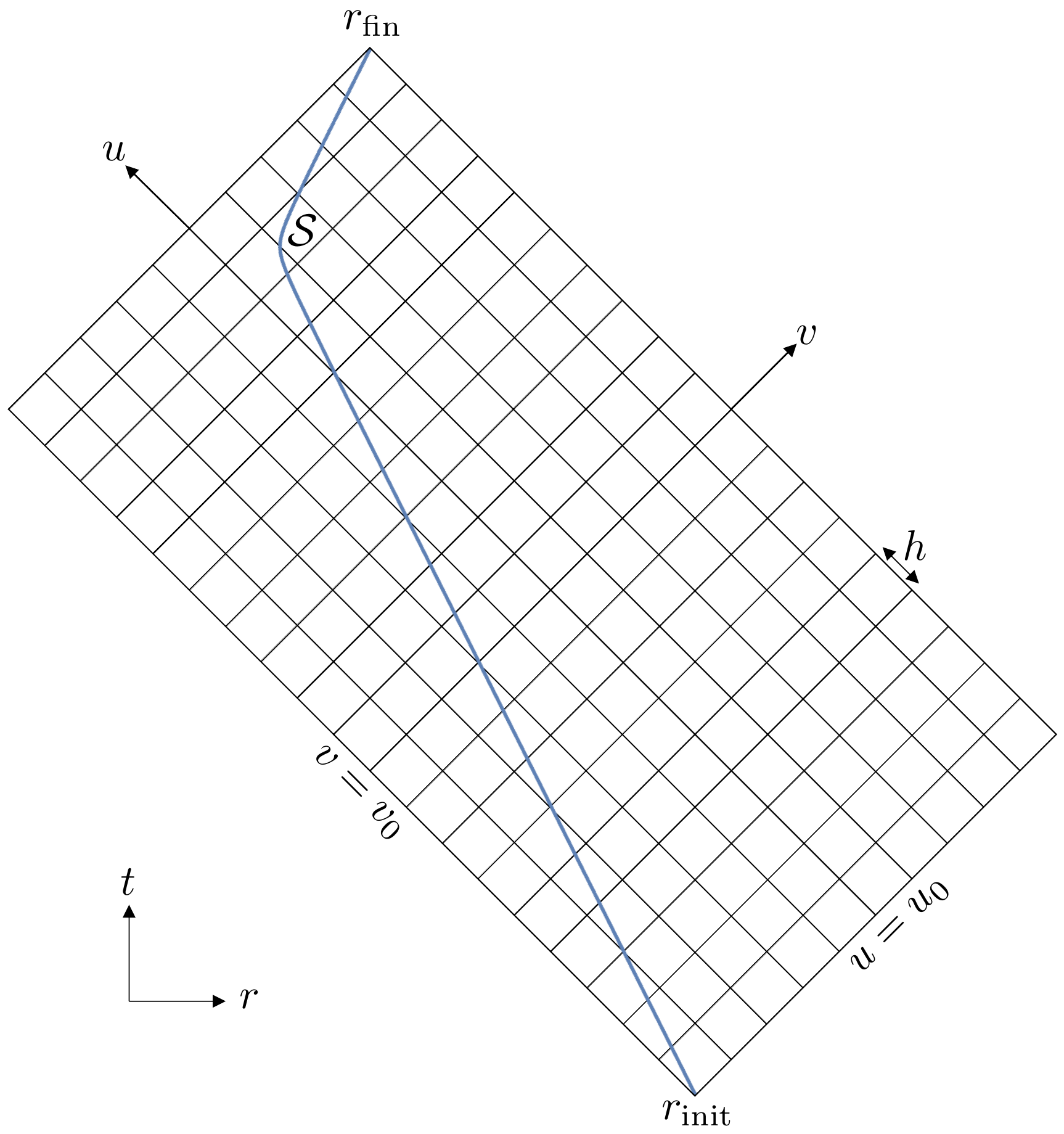}
\caption{Illustration of the 1+1D characteristic grid used in our numerical evolution of the scalar-field modes $\psi_{\ell m}(t,r)$. The curve $\cal S$ represents the scalar charge's fixed scattering geodesic worldline. Grid cells have uniform dimensions $\Delta v\times \Delta u=h\times h$, where $h$ is typically taken to be of order $\sim m_2/100$. Our finite-difference scheme has a quartic local convergence in $h$, and a demonstrated quadratic convergence globally. Finite-$h$ errors are controlled by running the code with a sequence of different $h$ values.  }
\label{uvGrid}
\end{figure}

We solve Eq.\ (\ref{eqn:SourcedFieldEquation}) in the time domain using a finite-difference scheme in null coordinates $u,v$, starting with characteristic initial data on two initial rays in the far past. Our finite difference scheme is described in detail in Appendix B of Ref.~\cite{Barack:2022pde}. The structure of the numerical grid is illustrated in Fig.\  \ref{uvGrid} here. It is set up so that the particle enters the numerical domain at the lower vertex, at $r=r_{\rm init}\gg m_2$, and leaves it (after being scattered) at the upper vertex, where $r=r_{\rm fin}\gg m_2$.
As initial data, we simply set $\psi_{\ell m}\equiv 0$ on the two initial rays, $u=u_0$ and $v=v_0$.  This unphysical set of initial data produces a burst of spurious (``junk'') radiation, which, however, dies off with a rapid power law in $t$.
The values of $r_{\rm init}$ and $r_{\rm fin}$ are chosen such that the junk data has sufficiently radiated away by $r=r_{\rm fin}$ on the {\em ingoing} leg and thus we have clean orbital data for all $r_p\leq r_{\rm fin}$, on both legs. The junk-contaminated data for $r_{\rm fin}<r_p\leq r_{\rm init}$ is discarded. 

We evolve the fields $\psi_{\ell m}$ for $-\ell \leq m \leq \ell$ with $0 \leq \ell \leq 15$, and record the value of the fields and their first derivatives along the scattering worldline. The results are then inputted into the (scalar-field version of the) mode-sum formula (\ref{mode-sum}) to obtain the full scalar-field SF. Section VII.D of Ref.~\cite{Barack:2022pde} provides the values of the regularization parameters featured in the mode-sum formula. High-order parameters, representing higher-order terms in the $1/\ell$ expansion down to $\ell^{-6})$, are known analytically and are incorporated in the mode sum to improve its convergence and control the error from the large-$\ell$ truncation.   

In the next step, the self-acceleration $F^\alpha$ is constructed via Eq.\ (\ref{eqn:SF_orth}), and then split into its conservative and dissipative pieces using Eq.\ (\ref{F_cons_diss}). Finally, we obtain the corresponding pieces of the SF correction to the scattering angle, $\delta\chi^{\rm cons}$ and $\delta\chi^{\rm diss}$, by numerically integrating the SF along the scattering geodesic using the appropriate versions of Eq.\ (\ref{deltaphi1_final_method2}). 

For the calculations reported here we took $r_{\rm fin}=1100m_2$, with $r_{\rm init}$ in the range $3750m_2\leq r_{\rm init}\leq 6600m_2$ as needed (larger $b$ or larger $v$ require larger $r_{\rm init}$). With these choices, a typical calculation of $\delta\chi$ for a single weak-field scattering orbit took $\sim 100$ days of CPU time split across 72 cores of the IRIDIS 5 cluster at the University of Southampton. The code is parallelized such that each independent $\psi_{\ell m}$ mode is computed simultaneously.

\subsection{Post-processing and error control}
\label{sec:PostProc}

The various sources of numerical error in our method are listed and analyzed in Sec.\ VIII B of Ref.~\cite{Barack:2022pde}. For the purpose of our current calculation, we have implemented several new post-processing algorithms to mitigate the two most dominant forms of error in the scattering-orbit case. We describe them in turn now. 

\subsubsection{Large-$r$ truncation}

By far the most dominant numerical error comes from the truncation of the orbital integral at a finite $r=r_{\rm fin}$. Increasing $r_{\rm fin}$ is highly punitive computationally since the runtime is proportional to $r_{\rm fin}^2$. In practice, the requirement of run times of $\text{day})$ per orbit restricted us to $r_{\rm fin}=1000m_2)$.  The orbital integral (\ref{deltaphi1_final_method2}) for $\delta\chi$, when truncated at $r_{\rm fin}$, converges like $\sim r^{-2}_{\rm fin}$, suggesting that neglecting the $r>r_{\rm fin}$ part of the integral produces a relative error of order $\sim(r_{\rm fin}/r_{\rm min})^{-2}$, where $r_{\rm min}$ is the periastron distance. In our runs $r_{\rm min}\sim 100m_2$ (cf.\ Tables \ref{Tablev50}-\ref{table4} below), and so the expected relative error from the truncation is $\sim (1000/100)^{-2}=1\%$. This level of error is more than we are willing to tolerate.  

To overcome this difficulty we fit an analytical polynomial model (a sum of powers of $1/r_p$) to a section of the large-$r_p$ SF data. We then use this to analytically extrapolate the SF to $r_p\to\infty$ on both legs of the orbit, which finally allows us to evaluate the integral in (\ref{deltaphi1_final_method2}) over the full scattering trajectory. 

We determine the residual error from applying this procedure by looking at the variation in the integral values when using models with differing polynomial orders and different spans of data used for fitting. The results are typically reliable to within $\sim 1\%$ of the tail contribution, leading to a residual error in $\delta\chi$ of order $\lesssim0.01\%$. This remains the dominant contribution to the error budget in our calculation, and therefore it sets the overall precision of our result for $\delta\chi$ at about one part in $10^{4}$.

It may be possible to obtain a large-$r_p$ approximation for the SF {\it analytically}, which would help reduce this source of error further. We plan to explore this route in future work. 

\subsubsection{Finite resolution}

The second most significant error is due to the finite resolution of the numerical grid used to evolve the fields $\psi_{\ell m}$. Decreasing $h$ (the stepping interval in $u$ and in $v$) is also highly punitive, since the runtime is proportional to $h^{-2}$. Given our resources, we were limited to grid dimensions $h\times h$ with $h$ not much smaller than $\sim m_2/100$.  At such resolutions, the finiteness of $h$ turns out to cause an error of $\sim 0.1\%$ in the final computed value of $\delta\chi$ (this error was estimated by varying over $h$). This would have become the dominant source of error if left unattended. 

Fortunately, it is possible to significantly reduce this source of error through a Richardson-type extrapolation, since the convergence properties of our finite-difference scheme are known.  Consider an exact value $\alpha_{\rm exact}$ described by a discretized numerical model $\alpha(h)$, such that
\begin{equation}
\alpha(h) = \alpha_{\rm exact} + C h^n + O\left( h^{n+1} \right),
\end{equation}
where $C$ is a constant and $h$ is the model resolution. Then, we can write 
\begin{equation}
\alpha_{\rm exact} = \frac{c^n \: \alpha(h/c) - \alpha(h)}{c^n - 1}   + O\left( h^{n+1} \right).
\end{equation} 
It follows that by evaluating the model with two different resolutions, $h$ and $h/c$, we can effectively increase the convergence rate of the model by one power of $h$. 

Our finite-difference algorithm has a global quadratic convergence, i.e.\ $n=2$ in the above expressions. By using the above extrapolation method with two different resolutions, we have obtained an effective {\em cubic} convergence. This reduces the error accumulated from the numerical integration such that the post-extrapolation finite-difference error is negligible compared to the error due to the large-$r_p$ analytic fit. 

\subsection{Sample SF results}

\begin{figure}[ht!]
\centering
\includegraphics[width=.8\linewidth]{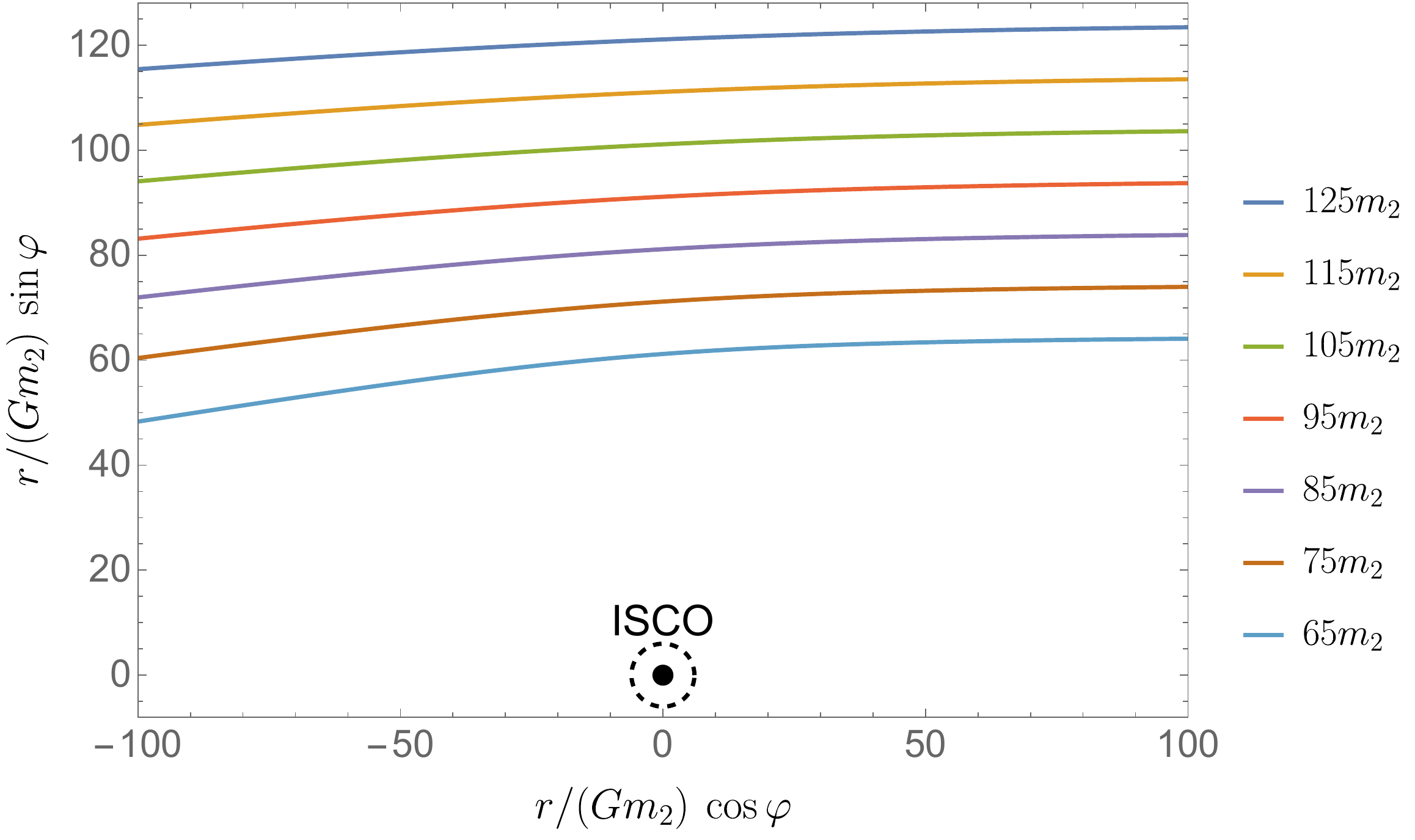}
\caption{
A subset of the sample of geodesic orbits used for our comparison. These scattering orbits all have $v=0.5$, with varying values of the impact parameter $b$ (indicated in the legend). The orbits are depicted in the equatorial plane of the Schwarzschild black hole (black disk at the bottom) using Cartesian coordinates constructed from the Schwarzschild coordinates $r,\varphi$. The location of the innermost stable circular orbit (ISCO) is shown in the dashed line for reference.}
\label{orbits}
\end{figure}

For this work we considered a large sample of weak-field scattering orbits, sampling the large-$b$ portion of the parameter space accessible to our SF code. Figure \ref{orbits} depicts a subset of our orbits, with $v=0.5$ (corresponding to $\sigma \simeq 1.1547$) and impact parameters in the range $60m_2 \leq b \leq 125m_2$. We have also sampled at fixed $v$ intervals in the range $0.5\leq v\leq 0.7$ for fixed $b=80m_2$, fixed $b=100m_2$, and fixed $b=125m_2$.

The full set of numerical results obtained for this work is displayed in Tables \ref{Tablev50}-\ref{table4}. For each of the orbits in our sample, the table shows the value of the periastron distance $r_{\rm min}$ and the corresponding geodesic scattering angle $\chi_0$, alongside the SF corrections (per $q_s$) $\delta\chi^{\rm cons}$ and $\delta\chi^{\rm diss}$ and the total correction $\delta\chi=\delta\chi^{\rm cons}+\delta\chi^{\rm diss}$. We also give estimated error bars. 

\begin{table}[!htb]
    \centering
    \begin{tabular}{ll|lrrr}
    \hline
\multicolumn{1}{c}{$b/m_2$} & \multicolumn{1}{c|}{$r_{\rm min}/m_2$} & \multicolumn{1}{c}{$\chi_0$} & \multicolumn{1}{c}{$\delta\chi$} & \multicolumn{1}{c}{$\delta\chi^{\rm cons}$} & \multicolumn{1}{c}{$\delta \chi^{\rm diss}$} \\ \hline
 60 & 55.9951 & 0.17905234 & $-2.5214(1) \times 10^{-4} $ & $-3.10227(7) \times 10^{-4} $ & $5.8086(3) \times 10^{-5} $ \\
 65 & 60.9958 & 0.16430723 & $-2.1266(1) \times 10^{-4} $ & $-2.5740(1) \times 10^{-4} $ & $4.4735(2) \times 10^{-5} $ \\
 70 & 65.9964 & 0.15181012 & $-1.8177(2) \times 10^{-4} $ & $-2.1695(1) \times 10^{-4} $ & $3.5181(2) \times 10^{-5} $ \\
 75 & 70.9969 & 0.14108256 & $-1.5712(2) \times 10^{-4} $ & $-1.8529(2) \times 10^{-4} $ & $2.8164(1) \times 10^{-5} $ \\
 80 & 75.9973 & 0.13177306 & $-1.3717(2) \times 10^{-4} $ & $-1.6006(2) \times 10^{-4} $ & $2.2896(1) \times 10^{-5} $ \\
 85 & 80.9976 & 0.12361754 & $-1.2077(2) \times 10^{-4} $ & $-1.3963(2) \times 10^{-4} $ & $1.88633(9) \times 10^{-5} $ \\
 90 & 85.9979 & 0.11641374 & $-1.0715(2) \times 10^{-4} $ & $-1.2287(2) \times 10^{-4} $ & $1.57248(7) \times 10^{-5} $ \\
 95 & 90.9981 & 0.11000409 & $-9.568(2) \times 10^{-5} $ & $-1.0892(2) \times 10^{-4} $ & $1.32453(6) \times 10^{-5} $ \\
 100 & 95.9983 & 0.10426402 & $-8.597(2) \times 10^{-5} $ & $-9.723(2) \times 10^{-5} $ & $1.12607(5) \times 10^{-5} $ \\
 105 & 100.998 & 0.09909373 & $-7.766(2) \times 10^{-5} $ & $-8.731(2) \times 10^{-5} $ & $9.6538(4) \times 10^{-6} $ \\
 110 & 105.999 & 0.09441235 & $-7.050(2) \times 10^{-5} $ & $-7.884(2) \times 10^{-5} $ & $8.3384(3) \times 10^{-6} $ \\
 115 & 110.999 & 0.09015361 & $-6.428(2) \times 10^{-5} $ & $-7.154(2) \times 10^{-5} $ & $7.2515(3) \times 10^{-6} $ \\
 120 & 115.999 & 0.08626271 & $-5.885(2) \times 10^{-5} $ & $-6.520(2) \times 10^{-5} $ & $6.3456(2) \times 10^{-6} $ \\
 125 & 120.999 & 0.08269395 & $-5.408(2) \times 10^{-5} $ & $-5.967(2) \times 10^{-5} $ & $5.5844(2) \times 10^{-6} $ \\ \hline
    \end{tabular}
    \caption{
    \label{Tablev50}
    SF numerical data for geodesic scattering orbits with $v=0.5$. For each orbit, the table displays the impact parameter $b$, the periastron distance $r_{{\rm min}}$, the geodesic scattering angle $\chi_0$ (per $q_s$), the total SF correction $\delta \chi$ (per $q_s$), and its conservative and dissipative pieces in separate. Parenthetical figures indicate the estimated numerical uncertainty in the last quoted decimal; e.g., $-2.5214(1) \times 10^{-4}$ means $-2.5214 \times 10^{-4} \pm 1\times 10^{-8}$.
    }
\end{table}

\begin{table}[!htb]
    \centering
    \begin{tabular}{ll|lrrr}
    \hline
\multicolumn{1}{c}{$v$} & \multicolumn{1}{c|}{$r_{\rm min}/m_2$} & \multicolumn{1}{c}{$\chi_0$} & \multicolumn{1}{c}{$\delta\chi$} & \multicolumn{1}{c}{$\delta\chi^{\rm cons}$} & \multicolumn{1}{c}{$\delta \chi^{\rm diss}$} \\ \hline
 0.525 & 76.3610 & 0.12186508 & $-1.3591(1) \times 10^{-4} $ & $-1.5655(1) \times 10^{-4} $ & $2.0638(1) \times 10^{-5} $ \\
 0.55 & 76.6777 & 0.11327894 & $-1.3477(1) \times 10^{-4} $ & $-1.5358(1) \times 10^{-4} $ & $1.8808(2) \times 10^{-5} $ \\
 0.575 & 76.9550 & 0.10578972 & $-1.3371(1) \times 10^{-4} $ & $-1.5103(1) \times 10^{-4} $ & $1.7316(2) \times 10^{-5} $ \\
 0.6 & 77.1992 & 0.09921831 & $-1.3274(1) \times 10^{-4} $ & $-1.4884(1) \times 10^{-4} $ & $1.6091(2) \times 10^{-5} $ \\
 0.625 & 77.4153 & 0.09342069 & $-1.3184(1) \times 10^{-4} $ & $-1.46933(9) \times 10^{-4} $ & $1.5089(2) \times 10^{-5} $ \\
 0.65 & 77.6074 & 0.08828002 & $-1.3102(1) \times 10^{-4} $ & $-1.45291(8) \times 10^{-4} $ & $1.4272(2) \times 10^{-5} $ \\
 0.675 & 77.7790 & 0.083700676 & $-1.30254(8) \times 10^{-4} $ & $-1.43867(5) \times 10^{-4} $ & $1.3614(2) \times 10^{-5} $ \\
 0.7 & 77.9328 & 0.079603825 & $-1.29542(4) \times 10^{-4} $ & $-1.42633(3) \times 10^{-4} $ & $1.3090(1) \times 10^{-5} $ \\ \hline
    \end{tabular}
    \caption{Same as in Table \ref{Tablev50}, for orbits with $b=80m_2$ and varying $v$.} \label{table2}
\end{table}

\begin{table}[!htb]
    \centering
    \begin{tabular}{ll|lrrr}
    \hline
\multicolumn{1}{c}{$v$} & \multicolumn{1}{c|}{$r_{\rm min}/m_2$} & \multicolumn{1}{c}{$\chi_0$} & \multicolumn{1}{c}{$\delta\chi$} & \multicolumn{1}{c}{$\delta\chi^{\rm cons}$} & \multicolumn{1}{c}{$\delta \chi^{\rm diss}$} \\ \hline
 0.525 & 96.3636 & 0.09644429 & $-8.532(2) \times 10^{-5} $ & $-9.551(2) \times 10^{-5} $ & $1.01836(5) \times 10^{-5} $ \\
 0.55 & 96.6813 & 0.08966728 & $-8.474(2) \times 10^{-5} $ & $-9.405(1) \times 10^{-5} $ & $9.3077(7) \times 10^{-6} $ \\
 0.575 & 96.9594 & 0.08375559 & $-8.420(1) \times 10^{-5} $ & $-9.279(1) \times 10^{-5} $ & $8.5923(9) \times 10^{-6} $ \\
 0.6 & 97.2041 & 0.07856798 & $-8.370(1) \times 10^{-5} $ & $-9.171(1) \times 10^{-5} $ & $8.004(1) \times 10^{-6} $ \\
 0.625 & 97.4205 & 0.07399089 & $-8.325(1) \times 10^{-5} $ & $-9.077(1) \times 10^{-5} $ & $7.523(1) \times 10^{-6} $ \\
 0.65 & 97.6128 & 0.06993216 & $-8.283(1) \times 10^{-5} $ & $-8.996(1) \times 10^{-5} $ & $7.132(1) \times 10^{-6} $ \\
 0.675 & 97.7845 & 0.06631638 & $-8.244(1) \times 10^{-5} $ & $-8.9256(9) \times 10^{-5} $ & $6.8172(8) \times 10^{-6} $ \\
 0.7 & 97.9383 & 0.063081378 & $-8.2079(7) \times 10^{-5} $ & $-8.8649(7) \times 10^{-5} $ & $6.5695(2) \times 10^{-6} $ \\ \hline
    \end{tabular}
    \caption{Same as in Table \ref{Tablev50}, for orbits with $b=100m_2$ and varying $v$.} \label{table3}
\end{table}

\begin{table}[!htb]
    \centering
    \begin{tabular}{ll|lrrr}
    \hline
\multicolumn{1}{c}{$v$} & \multicolumn{1}{c|}{$r_{\rm min}/m_2$} & \multicolumn{1}{c}{$\chi_0$} & \multicolumn{1}{c}{$\delta\chi$} & \multicolumn{1}{c}{$\delta\chi^{\rm cons}$} & \multicolumn{1}{c}{$\delta \chi^{\rm diss}$} \\ \hline
 0.525 & 121.366 & 0.07650367 & $-5.375(1) \times 10^{-5} $ & $-5.882(1) \times 10^{-5} $ & $5.0634(2) \times 10^{-6} $ \\
 0.55 & 121.684 & 0.07113856 & $-5.345(1) \times 10^{-5} $ & $-5.809(1) \times 10^{-5} $ & $4.6386(3) \times 10^{-6} $ \\
 0.575 & 121.963 & 0.06645828 & $-5.318(1) \times 10^{-5} $ & $-5.747(1) \times 10^{-5} $ & $4.2912(4) \times 10^{-6} $ \\
 0.6 & 122.208 & 0.06235107 & $-5.293(1) \times 10^{-5} $ & $-5.693(1) \times 10^{-5} $ & $4.0054(5) \times 10^{-6} $ \\
 0.625 & 122.425 & 0.05872706 & $-5.270(1) \times 10^{-5} $ & $-5.647(1) \times 10^{-5} $ & $3.7714(7) \times 10^{-6} $ \\
 0.65 & 122.617 & 0.05551334 & $-5.248(1) \times 10^{-5} $ & $-5.606(1) \times 10^{-5} $ & $3.5814(7) \times 10^{-6} $ \\
 0.675 & 122.789 & 0.05265023 & $-5.229(1) \times 10^{-5} $ & $-5.571(1) \times 10^{-5} $ & $3.4290(7) \times 10^{-6} $ \\
 0.7 & 122.943 & 0.05008854 & $-5.210(1) \times 10^{-5} $ & $-5.541(1) \times 10^{-5} $ & $3.3102(6) \times 10^{-6} $ \\ \hline
    \end{tabular}
    \caption{Same as in Table \ref{Tablev50}, for orbits with $b=125m_2$ and varying $v$.} \label{table4}
\end{table}

\section{Post-Minkowskian Results}
\label{sec:PM_results}

Following the amplitude-based methods of Refs.~\cite{Cheung:2018wkq, Bern:2019nnu, Bern:2019crd, Parra-Martinez:2020dzs, Bern:2021dqo, Bern:2021yeh} outlined in Sec.~\ref{sec:PM}, we compute the terms proportional to $q_s$ in the scattering angle.  We compute contributions through ${\cal O}(q_s G^3)$, keeping only those terms that are of first order in the SF expansion, in order to match the self-force calculation described in Sec.~\ref{sec:NumericalSF}.

\subsection{Conservative dynamics}

\subsubsection{Conservative amplitudes}
\label{subsec:cons_pm_results}
We start by decomposing the scattering amplitude into a geodesic piece and a first-order SF contribution (at fixed $v$ and $b$):
\begin{equation}
    \calM = \calM^{\probe} + q_{s}\,\delta \calM +\calO(q_s^2, q_m) .
\end{equation}
In turn, $\delta \calM$ is decomposed into minimal-coupling and tidal contributions:
\begin{equation}
    \delta \calM = \delta \calM^{\rm min} + \delta \calM^{\rm tidal}.
\end{equation}
Computing minimal-coupling amplitudes as described in \Sec{subsec:PM_cons} up to fourth PM order, we find
\begin{align}
\label{eq:amp_min_1}
 \delta \mathcal M^{\rm min}_{1}(q) = {}&
0
,
\\
 \label{eq:amp_min_2}
\delta \mathcal M^{\rm min}_{2}(q)= {}&
-2\pi^2 G m_1^2 m_2^3\frac{1}{|q|}(\sigma^2-1)
,
\\
\label{eq:amp_min_3}
 \delta \mathcal M^{\rm min}_{3}(q)= {}&
 \frac{8\pi}{3}G^2 m_1^2 m_2^4\, \log\left(\frac{\bar\mu^2}{|q|^2}\right)\,
 \sigma(1+2\sigma^2) 
 + \mathrm{iteration} 
,
\\
\label{eq:amp_min_4}
\delta \mathcal M^{\rm min}_{4}(q)={}&
G^3\pi^2m_1^2m_2^5 |q| 
\left(\frac{4^{1/3}\bar \mu^2}{|q|^2}\right)^{3\epsilon} \! 
\left\{
    \left[
        4\mathcal{M}_4^{\mathrm{t}} \log\left(\tfrac{\sqrt{\sigma^2-1}}{2}\right)
        + \mathcal{M}_4^{\pi^2}
        +\mathcal{M}_4^{\mathrm{rem}}
    \right]
- \frac{(\sigma^2-1)} {\epsilon} \right\} \nn\\
&  \hskip 4 cm \null + \mathrm{iteration} ,
\end{align}
where we have suppressed iteration terms of lower-order quantities, and $\bar \mu^2=\mu^2e^{\gamma_E} /(4\pi)$ denotes the $\overline{\text{MS}}$ scheme dimensional regularization scale and $\gamma_\text{E}$ is the Euler-Mascheroni constant. Notice that the $\operatorname{arccosh}(\sigma)$ function found in Ref.~\cite{Bern:2019nnu} is absent in the expression for $\delta \calM_3(q)$, similar to scalar electrodynamics \cite{Saketh:2021sri,Bern:2021xze}. In $\delta \calM^{\rm min}_4(q)$  we introduced the short-hand notion
\begin{align} \label{eq:M4Defintions}
\begin{split}
{\cal M}_4^{\rm t} &= 
	r_1 + r_2 \log \left(\tfrac{\sigma +1}{2}\right) +  r_3\frac{  {\rm arccosh} (\sigma) }{\sqrt{\sigma^2-1}}
	, \\
 {\cal M}_4^{\pi^2} &= 
	  r_5\, K \big(\tfrac{\sigma -1}{\sigma +1}\big) E\big(\tfrac{\sigma -1}{\sigma +1}\big) 
                    + r_6\, K^2\big(\tfrac{\sigma -1}{\sigma +1}\big)
                + r_7\, E^2\big(\tfrac{\sigma -1}{\sigma +1}\big)  
 , \\
	{\cal M}_4^{\rm rem} &=  
	r_8
	+ r_9 \log \left(\tfrac{\sigma +1}{2}\right) 
	+ r_{10}  \tfrac{ {\rm arccosh}(\sigma )   }{\sqrt{\sigma^2-1}}
	 + r_{12}\log^2 \left(\tfrac{\sigma +1}{2}\right)
	+ r_{14} \, \tfrac{ {\rm arccosh}^2(\sigma )}{\sigma^2-1}
	.
\end{split}
\end{align}
The $r_i$ are rational functions of $\sigma$; they are detailed in Eq.~\eqref{table:functions} and their numbering follows that of the functions in Ref.~\cite{Bern:2021yeh}. In Eq.~\eqref{eq:M4Defintions} $E$ and $K$ are the complete elliptic integrals of the first and second kinds that appear in the analogous results in General Relativity. Compared to the corresponding result in General Relativity, it is also noteworthy that the higher-weight polylogarithmic functions do not appear in the scalar model and six rational coefficients vanish. 
\begin{align}
\begin{array}{rlcrl}
r_{1}  &=
\frac{3 \sigma ^4}{2}-\sigma ^3+\frac{3 \sigma ^2}{4}+\sigma -\frac{19}{12}
,
&
&
r_{2}  &=
-\frac{3}{2} \left(\sigma ^2-1\right)^2
\! ,
\cr
r_{3}  &=
\frac{3}{4} \sigma {} \left(\sigma ^2-1\right) \left(2 \sigma ^2-3\right)
,
&&
r_5 & = 
-\frac{1}{2}\frac{100 \sigma ^2+177 \sigma +79}{\sigma ^2-1}
,
\cr
r_6 & = 
\frac{95 \sigma +82}{2 \left(\sigma ^2-1\right)}
,
&&
r_7 & = 
\frac{100 \sigma ^2+79}{4 (\sigma -1)} 
,
\cr
r_8 & = 
\frac{18 \sigma ^6+252 \sigma
   ^5-347 \sigma ^4-216 \sigma
   ^3+711 \sigma ^2-348 \sigma
   -14}{12 \left(\sigma
   ^2-1\right)}
,
&&
r_9 & = 
\frac{1}{2} \left(3 \sigma ^4+8 \sigma ^3-22 \sigma ^2-8 \sigma +27\right)
,
\cr
r_{10} & = 
-\frac{\sigma  \left(2 \sigma ^2-3\right) \left(45 \sigma ^4-29\right)}{12 \left(\sigma ^2-1\right)}
,
&&
r_{12} & = 
3 \left(\sigma ^2-1\right)^2
\! ,
\cr
r_{14} & =
-\frac{3}{4} \sigma ^2 \left(3-2 \sigma ^2\right)^2 
\! ,
&&
r_4 & = r_{11} = r_{13} = r_{15} = r_{16} = r_{17} = 
0
.
\end{array}
\label{table:functions}
\end{align}

\subsubsection{Finite-size Effects}

In comparison to General Relativity at the loop orders considered here, the scalar model shows a new nontrivial feature. The UV divergence in Eq.~\eqref{eq:amp_min_4} reminds us that, in addition to the minimal-coupling contributions of \Eq{eq:model_action_minimal}, we need to consistently take into account the nonminimal tidal operators of \Eq{eq:tidal_operators}. The nonminimal tidal action of Eq.~\eqref{eq:tidal_operators} gives rise to the following tree-level amplitude (all out-going momentum convention):
\begin{align}
\label{eq:tidal_tree}
    \calM^{{\rm tidal}}(\chi,\phi_2,\phi_2,\chi)=
    \hspace{-2.3cm}\vcenter{\hbox{\scalebox{.9}{\ctTree}}}
    \hspace{-2cm}
    =
   -16 \pi G^3 m_2^4 \left[c_1 (\ell_1\cdot \ell_4) + (c_1{+}c^{{\rm bare}}_2) (\ell_1\cdot u_2)(\ell_4\cdot u_2)\right]
   + {\rm quantum}.
\end{align}
In writing Eq.~\eqref{eq:tidal_tree} we have already expanded the tree amplitude in the classical limit relevant for the problem of interest, where we take the momenta of the scalar field $\psi$ to be of order $\ell_{1},\ell_4\ll m_2 u_2$.  From the scaling analysis described in Section \ref{subsec:EFT_and_tidal}, the tidal operators add to the minimal coupling amplitude at $\calO(G^3 q_s)$ through the operator insertion into a one-loop diagram:
\begin{align}
\label{eq:3PM_ct}
    \delta \mathcal M^{{\rm tidal}}_{4}(q)=
    \hspace{-1.5cm}\vcenter{\hbox{\scalebox{.9}{\ctOneLoopV}}}
    \hspace{-2cm}= 
    G^3\, 4 \pi^2\, m_1^2 m_2^5 |q|
    \left(16\frac{\bar \mu^2}{|q|^2}\right)^\epsilon  {}
    \left[c_1 - (c_1{+}c^{{\rm bare}}_2) \frac{(\sigma^2-1)}{4(1-\epsilon)}\right]
    ,
\end{align}
where we performed the tensor reduction and evaluated the resulting scalar triangle Feynman integral in the soft limit. Requiring the absence of a UV divergence in the full EFT amplitude yields a constraint on the divergent part of the bare Wilson coefficient $c^{{\rm bare}}_2$ as can be seen by comparing Eqs.~\eqref{eq:amp_min_4} and \eqref{eq:3PM_ct}:
\begin{equation}
c^{{\rm bare}}_2 = \left[-\frac{1}{\epsilon} + c_2(\bar \mu) \right]. 
\end{equation}
A standard effect in quantum field theory is the ``running'' of the finite part of the Wilson coefficient, $c_2(\bar \mu)$,  with dimensional regularization scale $\bar\mu$. On a technical level, this arises from the fact that the $\mathcal{O}(\epsilon)$ powers of the $|q|^2$ scaling do not match between the minimal three-loop amplitude in Eq.~\eqref{eq:amp_min_4} and the one-loop amplitude with counter-term insertion in Eq.~\eqref{eq:3PM_ct}. Conceptually, it means that the coupling depends on the energy scale of observation. 
As an additional consistency check, the same divergent value for $c^{{\rm bare}}_2$ also cancels the divergence in the two-loop $\phi \phi \psi \psi$ amplitude. 
Finally, we combine the part of the amplitude generated from the tidal contributions and the minimal amplitude, to give
\begin{align}
\label{eq:M4PM_full}
\begin{split}
\delta \calM_4 & = \delta \calM^{{\rm min}}_4 + \delta \calM^{{\rm tidal}}_4 
\\
& = 
G^3\pi^2m_1^2m_2^5\, |q|  \bigg\{
\left[4\mathcal{M}_4^{\mathrm{t}}\log\left(\tfrac{\sqrt{\sigma^2-1}}{2}\right)+\mathcal{M}_4^{\pi^2}+\mathcal{M}_4^{\mathrm{rem}}\right] \\
&\hspace{1cm}
-4 c_1-(c_1 {+} c_2(\bar\mu)-1) \left(\sigma ^2-1\right)
+2 \left(\sigma ^2-1\right) \log \left(\frac{\bar \mu^2}{2|q|^2}\right)
\bigg\}
\\
& \hspace{1cm} + \mathrm{iteration}.
\end{split}
\end{align}
We have checked and confirmed that the coefficient of $\log(\sqrt{\sigma^2-1}/2)$ in Eq.~\eqref{eq:M4PM_full} is proportional to the energy loss at order $\calO(G^2q_s)$ in the scalar model as expected~\cite{Bini:2017wfr, Blanchet:2019rjs, Bini:2020hmy, Bern:2021dqo}. We will present results for the energy loss later, when we describe the radiative dynamics of the scalar-field model. As one can see, canceling the UV divergence leads to a finite logarithm that depends on the scale $\bar{\mu}$. The full result is of course invariant under this choice. Therefore the coefficient $c_2(\bar{\mu})$ must cancel this scale dependence through the well-known renormalization-group equation, ${d c_2(\bar{\mu})}/{d\log \bar{\mu}} = 4$. This equation allows us to change the choice of scale from one to another without changing the physics. For practical convenience, we can pick $\bar{\mu}$ to be the inverse of the Schwarzschild radius of the heavy black hole. The coefficient $c_2$ we fit later with SF numerical result is measure at this scale.

\subsubsection{Radial action}

We decompose the radial action for the full theory into a purely gravitational piece $I_r$ and a scalar-field SF correction $\delta I_r$ (at fixed $v$ and $b$), in the form
\begin{equation}
    I_r(b)=I_r^{\probe}(b)+q_{s}\delta I_r(b) + \mathcal{O}(q_s^2,q_m).
\end{equation}
Computing the amplitudes as described in Section~\ref{subsec:PM_cons} and using the amplitude-action relation detailed in Eq.~\eqref{eq:amp_rad_act}, we find, up to third PM order,  
\begin{align}
\label{eq:radial_actiion_results}
 \delta I_{r,1}(b) = {}& 
 0
\,,
 \\[4pt]
 \delta I_{r,2}(b)= {}&
 -\frac{\pi G m_1 m_2^2}{4 b}\sqrt{\sigma^2-1}
 \,,
 \\[4pt]
 \delta I_{r,3}(b)= {}&
 - \frac{2 G^2 m_1 m_2^3}{3 b^2}\frac{\sigma (1+2\sigma^2)}{\sqrt{\sigma^2-1}}
 \,,
 \\[4pt]
 \begin{split}
\delta I_{r,4}(b)={}&
 \frac{\pi\, G^3 m_1 m_2^4}{8 b^3 \sqrt{\sigma^2-1}} \bigg\{
 -\left[
    4\mathcal{M}_4^{\mathrm{t}} \log\left(\tfrac{\sqrt{\sigma^2-1}}{2}\right)
    +\mathcal{M}_4^{\pi^2}
    +\mathcal{M}_4^{\mathrm{rem}}
  \right]
  \\
  & \hspace{2.5cm}
  + c_1 (\sigma^2-5) + \left(
  c_2(\bar\mu)
  -9 
  + 2 \log\left[2b^2 e^{2\gamma_{\rm E}} \bar \mu^2\right] 
  \right) (\sigma^2-1) 
 \bigg\}
 \,,
\end{split}
\end{align}
where we use the notation of Eq.~\eqref{eq:M4Defintions} and the $r_i$ detailed in Eq.~\eqref{table:functions}. These results have been verified up to 3PM by an independent worldline EFT calculation \cite{EMREFT}.

\subsubsection{Scattering angles}

Equipped with the explicit results for the SF corrections to radial action in \Eq{eq:radial_actiion_results}, we obtain the conservative SF corrections to the scattering angle up to 4PM order:
\begin{align}
 \delta\chi^{{\rm cons}} = -\frac{1}{m_1 \sqrt{\sigma^2-1}} \frac{\partial \delta I_{r}(b)}{\partial b} ,
\end{align}
with
\begin{align}
\delta\chi_{1}^{\rm cons} &= 
0 
, 
\\
\delta\chi_{2}^{\rm cons} & = 
-\frac{\pi}{4} G\frac{m_2^2}{ b^2}
, \label{eqn:2PMCons}
\\
\delta\chi_{3}^{\rm cons} & = 
-\frac{4}{3}G^2\frac{\sigma (1+2\sigma^2)}{(\sigma^2-1)}\frac{m_2^3}{ b^3}
, \label{eqn:3PMCons}
\\
\delta\chi_{4}^{\rm cons} & = 
\pi G^3\frac{3 m_2^4}{8 (\sigma^2-1) b^4}\Bigg\{
-\left[
    4\mathcal{M}_{4}^{\mathrm{t}} \log\left(\tfrac{\sqrt{\sigma^2-1}}{2}\right)
    +\mathcal{M}_4^{\pi^2}
    +\mathcal{M}_4^{\mathrm{rem}}
\right]\nonumber
\\
& \hspace{1.5cm}
  + c_1 (\sigma^2-5) 
  + \left(
  c_2(\bar\mu)
  -\frac{31}{3}
  + 2 \log\left[2 b^2 e^{2\gamma_{\rm E}} \bar \mu^2\right] 
  \right) (\sigma^2-1) 
\Bigg\}
.
\label{eqn:4PMCons}
\end{align}
The 2PM contributions have already been worked out in Ref.~\cite{Gralla:2021qaf} and compared to self-force results in Ref.~\cite{Barack:2022pde}.  In our setup these contributions are described by the single Feynman diagram in Fig.~\ref{fig:OneLoopDiag_scalar} \subref{fig:2pm_eg_diag}.

As a non-trivial check of our results we have verified that the angles satisfy the predicted iteration structure when expanded in the limit $ v \to 0$. In particular, we have \cite{Bern:2019crd}
\begin{align}
\label{eq:Iteration3}
\delta\chi_3^{\rm cons}={}&
        \frac{2 G^2\delta P_3}{m_1^2(\sigma^2-1)b^3}
        +2\frac{\delta \chi_2^{\rm cons}\chi_1^{(0)}}{\pi}
        =\mathcal{O}(v^{-2}), 
\\
\begin{split}
\label{eq:Iteration4}
\delta\chi_4^{\rm cons}={}&
        \frac{3 \pi}{4} \frac{G^3 \delta P_4}{m_1^2(\sigma^2-1)b^4}
        {+}\frac{3\pi}{8}\delta\chi_3^{\rm cons}\chi_1^{(0)}
        {-}\frac{3}{4}\delta\chi_2^{\rm cons}\left(\chi_1^{(0)}\right)^2
        {+}\frac{3}{\pi}\delta\chi_2^{\rm cons}\chi_2^{(0)}
        \\
        ={}&-\frac{9\pi}{4}\frac{G^3m_2^4}{(\sigma^2-1)^2 b^4}+\calO(v^{-2}),
\end{split}
\end{align}
where $p_\infty=m_1\sqrt{\sigma^2-1}$ and the $\delta P_k=\calO(v^0)$ is defined through the expansion of the radial momentum $p_r$  \cite{Bern:2019crd},
\begin{equation}
    p^2_r(r)=p_\infty^2-\frac{J^2}{r^2}+\sum_{k} \left(P^{(0)}_k+q_s^2\delta P_k\right)\frac{1}{r^k} +\calO(q_s^2,q_m).
\end{equation}
In Eqs.~\eqref{eq:Iteration3}--\eqref{eq:Iteration4} we have used the values of the geodesic scattering angle, expanded in $G$:
\begin{equation}
    \chi^{(0)} =\chi^{(0)}_1+\chi^{(0)}_2 +\calO(G^3)= \frac{G m_2}{ b}2\frac{2\sigma^2-1}{\sigma^2-1}+\frac{G^2m_2^2}{ b^2}\frac{3\pi}{4}\frac{5\sigma^2-1}{\sigma^2-1}+\calO(G^3).
    \label{eq:0SFangle}
\end{equation}

\newpage
\subsection{Radiative Dynamics}
As explained in \Sec{subsec:PM_rad}, we need the radiated angular momentum and energy at 2PM and 3PM orders for the 4PM dissipative scattering angle.
At 2PM, only the radiated angular momentum is non-vanishing due to the waveform in the zero-frequency limit.
We obtain the scalar current using Eq.~\eqref{eq:softlimit} with the geodesic scattering angle at 1PM in \Eq{eq:0SFangle}.
Plugging the current into into Eq.~\eqref{eq:EJrad_Genformula}, we obtain the radiated angular momentum at 2PM:
\begin{align}
	\delta J_{\rm rad,2} &= 
\frac{2m_1 m_2}{3}\left(\frac{G m_2}{ b}\right) (2\sigma^2-1)
.
\end{align}
Together with $\delta E_{\rm rad,2}=0$, the linear response formula yields the 3PM dissipative angle
\begin{align}
    \delta \chi^{\rm diss}_3 
	&= 
 \frac{2m_2}{3  b} \left(\frac{G m_2}{ b}\right)^2
	\frac{(2\sigma^2-1)^2}{(\sigma^2-1)^{3/2}}
.
\label{eq:chi_dis_3PM}
\end{align}
One can see that the full scattering angle at 3PM has a milder high-energy behavior than the conservative or dissipative ones alone, 
\begin{equation}
    \delta \chi^{\rm cons}_3+\delta \chi^{\rm diss}_3=0+\calO(\sigma^{-1}).
\end{equation}
At 3PM, we first calculate the tree-level amplitude of $\mathcal{M}(\phi_1,\phi_2,\phi_1,\phi_2,\psi)$ and then obtain the scalar waveform from \Eq{eq:waveformKMOC}.
Plugging the waveform into \Eq{eq:EJrad_Genformula} yields the radiated energy and angular momentum following the procedure in Ref.~\cite{Herrmann:2021lqe,Herrmann:2021tct}:
\begin{align}
    \delta E_{\rm rad,3} = 
    \frac{\pi m_1 m_2}{ b}\left(\frac{G m_2}{ b}\right)^2
    \bigg[&
    \frac{3}{16}\sigma {} (2 \sigma^2-3)
    \,{\rm arccosh}(\sigma )
    -\frac{3}{8}(\sigma^2-1)^{3/2}
    \,\log \left(\tfrac{\sigma +1}{2}\right)
    \nn \\
    &+\frac{18 \sigma^4- 12 \sigma^3+9 \sigma^2+ 12 \sigma -19}{48 \sqrt{\sigma^2-1}}
   \bigg]
.
\end{align}
As mentioned above, we see that the coefficient of a particular logarithm in the scattering amplitude \eqref{eq:amp_min_1}-\eqref{eq:amp_min_4}
is proportional to the energy loss 
\begin{equation}
     \delta E_{\rm rad,3}=\frac{G^2m_1m_2^3\pi}{4 b^3\sqrt{\sigma^2-1}}\mathcal{M}_4^{\mathrm{t}}.
\end{equation}
The radiated angular momentum at this order receives contributions from both the finite frequency and the zero-frequency limit.
The contribution from the latter is similar to the 2PM radiated angular momentum.
Using the procedure in Ref.~\cite{Manohar:2022dea}, the full answer reads
\begin{align}
    \delta J_{\rm rad,3} =
	\pi m_1 m_2\left(\frac{G m_2}{ b}\right)^2
    \bigg[&
    -\frac{3 \sigma^2 (2\sigma^2-3)}{8 \sqrt{\sigma^2-1}}\,{\rm arccosh}(\sigma )
    +\frac{3}{4}(\sigma^2-1)\sigma \log \,\left(\tfrac{\sigma +1}{2}\right)
    \nn \\
    &+\frac{9 \sigma^5 - 3 \sigma^4 - 3 \sigma^3 + 6 \sigma^2 + 2 \sigma - 3}{24 (\sigma^2-1)}
    \bigg]
.
\end{align}
Given both the radiated energy and angular momentum and the scattering angle in the geodesic limit, 
we obtain the scattering angle using the linear response formula. It reads
\begin{align} \label{eq:chi_dis_4PM}
    \delta \chi^{\rm diss}_4 =
	\frac{\pi m_2}{ b}\left(\frac{G m_2}{ b}\right)^3
    \bigg[&
    -\frac{3 \sigma^2 (12 \sigma^4 - 28 \sigma^2+15)}{16 (\sigma^2-1)^2}
    \,{\rm arccosh}(\sigma )
    +\frac{3 \sigma(6\sigma^2-5)}{8\sqrt{\sigma^2-1}} \log \,\left(\tfrac{\sigma +1}{2}\right)
    \nn \\
    &+\frac{252 \sigma^6+ 6 \sigma^5 - 438 \sigma^4 + 79 \sigma^3 + 204 \sigma^2 - 61 \sigma -18}{48 (\sigma^2-1)^{\frac{5}{2}}}
    \bigg]
.
\end{align}
%

\section{SF-PM comparison and fitting}
\label{sec:Comparison}

\begin{figure}[h!]
\centering
\includegraphics[width=\linewidth]{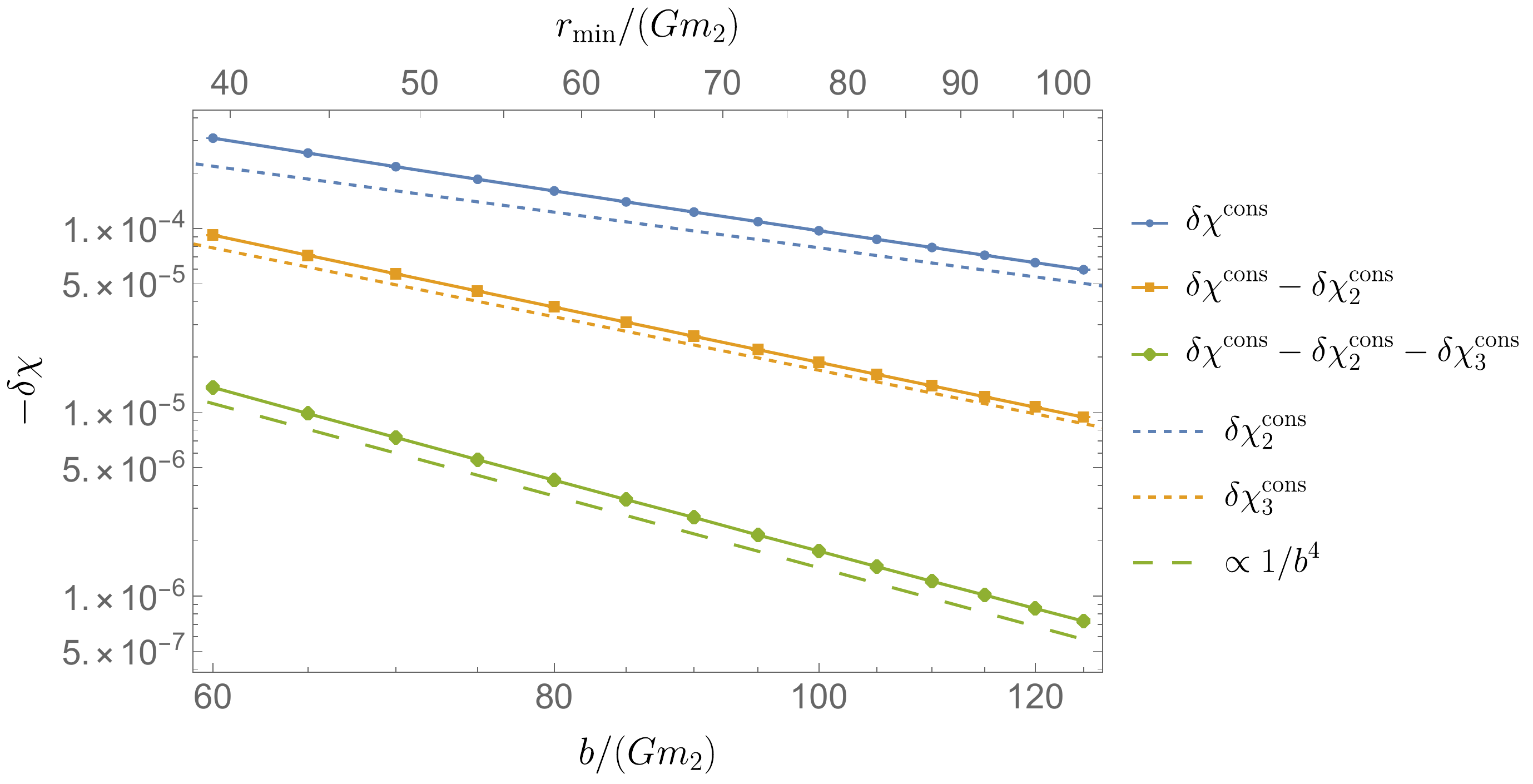}
\caption{
The conservative SF correction to the scattering angle (per $q_s$) for the values shown in Table \ref{Tablev50}, with $v=0.5$. The results are shown as a function of the impact parameter (lower scale) and corresponding minimal separation (upper scale). Error bars on the numerical data are too small to be seen on this scale. The upper curve (data points connected by a solid blue line) represents the numerical SF data for $\delta\chi^{\rm cons}$, and the adjacent dashed blue line shows the leading-order analytical PM term, $\delta\chi^{\rm cons}_2$. The solid and dashed orange curves in the middle show, respectively, the difference $\delta\chi^{\rm cons}-\delta\chi^{\rm cons}_2$ and the next-to-leading-order term $\delta\chi^{\rm cons}_3$. Finally, the solid green curve at the bottom displays the difference $\delta\chi^{\rm cons}-\delta\chi^{\rm cons}_2-\delta\chi^{\rm cons}_3$. We do not have the $\delta\chi^{\rm cons}_4$ term analytically for comparison, so instead we present (long dashed green line) a reference line $\sim b^{-4}$ with an arbitrary amplitude, showing the numerical data agree well with the analytical PM expressions through 3PM order. 
}
\label{v50ConsSub}
\end{figure}

\begin{figure}[h!]
\centering
\includegraphics[width=\linewidth]{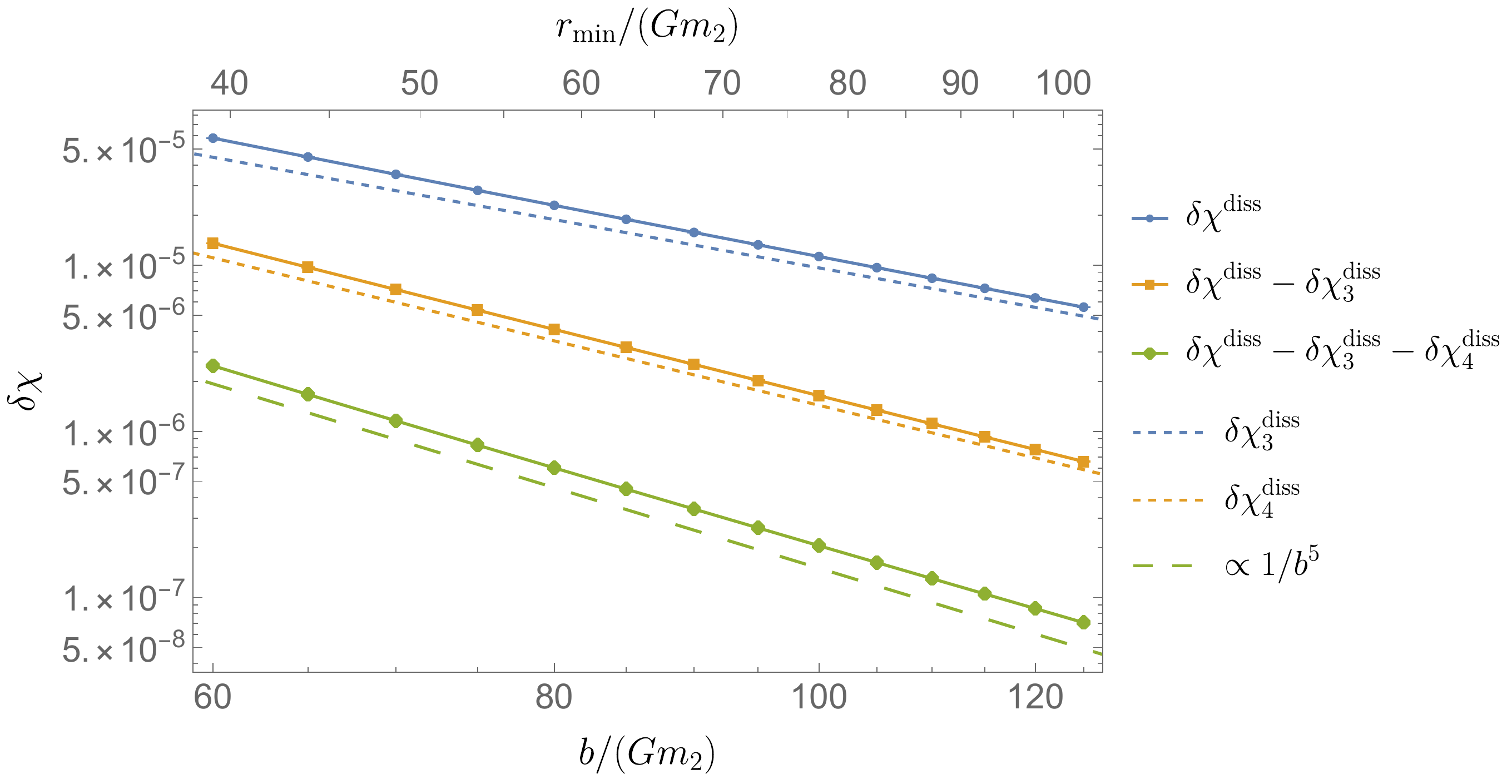}
\caption{
Similar to Fig.\ \ref{v50ConsSub} but for the dissipative correction $\delta\chi^{\rm diss}$. A good agreement between the SF data and the analytical PM results is manifest through 4PM order.}
\label{v50DissSub}
\end{figure}

We now turn to the comparison between the numerical SF results and the analytical PM expressions for the conservative and dissipative SF corrections to the scattering angle.  Figures \ref{v50ConsSub} and \ref{v50DissSub} show the behavior of $\delta\chi^{\rm cons}$ and $\delta\chi^{\rm diss}$ (respectively) as functions of impact parameter $b$ (lower scale) and the corresponding minimal separation $r_{\rm min}$ (upper scale), at a fixed $v=0.5$. In each plot, the upper (blue) data points show the base numerical values for $\delta\chi$, as compared to the leading-order (LO) analytical PM terms (dashed blue curve), i.e.\ $\delta\chi_2^{\rm cons}$ and $\delta\chi_3^{\rm diss}$. A close agreement is evident. The numerical-data curve appears to fall off slightly faster than $b^{-2}$ (conservative) or $b^{-3}$ (dissipative). This may be attributed to the fact that next-to-leading-order (NLO) PM terms have significant contribution in this part of the parameter space.

Removing the LO contribution from the numerical data (orange) gives a slope that closely resembles the analytical NLO term. In particular, the slope of the difference appears to be very close to $b^{-3}$ (conservative) or $b^{-4}$ (dissipative), confirming the good agreement between the data and LO PM term.

Finally, subtracting both LO and NLO PM terms from the numerical data (green) we find a fall-off consistent with $\sim\! b^{-4}$ (conservative) and $\sim\! b^{-5}$ (dissipative), confirming the good agreement between the data and the NLO PM term.

Notably, even after subtracting the two leading PM terms, the residual still shows a clear, smooth, power-law fall-off with $b$, suggesting our data has sufficient accuracy to encode information about subsequent terms in the PM expansion. In what follows we explore this possibility. By fitting the residual data to a model, we will attempt to extract information about the yet-unknown pieces $\delta\chi_{>3}^{\rm cons}$, and $\delta\chi_{>4}^{\rm diss}$.

Our goal here is not a precise determination of the high-order PM terms (this we will not do), but rather to provide a proof of concept, explore the limits of what can be done with the current data, and illustrate how such a fitting procedure might work in practice once a more accurate and exhaustive dataset is at hand. 
For our fitting we employ \texttt{Mathematica}'s function \texttt{NonlinearModelFit}, weighting each data point by the inverse of the square of the estimated numerical error. 

\subsection{Conservative sector: fitting for $\delta\chi^{\rm cons}_{>3}$}

Let us write
\begin{equation}
\label{model_cons}
\delta\chi^{\rm cons}(v,b) = \sum_{n=2}^{\bar n} \frac{a_n(v)}{b^n} + O\left(b^{-\bar n-1}\right)\,,
\end{equation}
where, for simplicity, we ignore (for now) logarithmic terms that occur at 4PM order and higher orders. The coefficients $a_2\equiv\delta\chi^{\rm cons}_2$ and $a_3\equiv\delta\chi^{\rm cons}_3$ are fully known: they are given in Eqs.\ (\ref{eqn:2PMCons}) and (\ref{eqn:3PMCons}), respectively. The coefficient $a_4\equiv\delta\chi^{\rm cons}_4$ is known up to two constant coefficients, $c_1$ and $c_2$---see Eq.~(\ref{eqn:4PMCons}). The coefficients $a_{>4}(v)$ are completely unknown.  The truncation order $\bar n$ can be used as a control parameter to monitor the quality of the fit. One hopes to find a range of $\bar n$ for which the fitted values of the coefficients $a_n$ are relatively stable under variation of $\bar n$. We expect the quality of the fit to deteriorate if $\bar n$ is taken too large (because of degeneracy) or too low (to the extent that the model fails to capture important higher-order terms). 

As a test of this procedure, we first use the data to fit for the {\it known} PM terms, and then compare with the known results. An example is shown in the first 4 rows of Table \ref{Table:fit_cons}, where no analytical knowledge is assumed, and we try fit the data (at fixed $v=0.5$) to the model (\ref{model_cons}) with varying values of $\bar n$ ($\bar n=2$ in the first row, $\bar n=3$ in the second row, etc.). We see that the fitted value of $a_2$ settles to within a fraction of a percent of its analytically predicted value, $a_2 \sim\! -0.785398$. The fitted value of the NLO coefficient, $a_3$, is less stable, but still varies within a few percents of the known analytical value, $a_3 \sim\! -16.9356$. The agreement gets better when the fit is performed with the analytical value of $a_2$ fixed in the model (fifth to seventh rows). These experiments provide further reassurance about the validity of our analytical results at 2PM and 3PM.

\begin{table}[h]
\begin{center}
\begin{tabular}{l|l|r|r}
\hline
\multicolumn{1}{c|}{$a_2$}            & \multicolumn{1}{c|}{$a_3$}           & \multicolumn{1}{c|}{$a_4$}       & \multicolumn{1}{c}{$a_5$}          \\ \hline
$-1.0886$     &       \multicolumn{1}{c|}{--}          &      \multicolumn{1}{c|}{--}      &          \multicolumn{1}{c}{--}       \\ 
$-0.7535$     &       $-21.77$          &      \multicolumn{1}{c|}{--}       &          \multicolumn{1}{c}{--}       \\ 
$-0.7899$     &       $-16.17$          &      $-206.5$       &          \multicolumn{1}{c}{--}       \\ 
$-0.7803$     &       $-18.49$          &      $-25.0$       &          $-4620$       \\ 
${\bf -0.785398}$     &       $-19.18$          &      \multicolumn{1}{c|}{--}       &          \multicolumn{1}{c}{--}       \\ 
${\bf -0.785398}$     &       $-16.93$          &      $-176.2$       &          \multicolumn{1}{c}{--}       \\ 
${\bf -0.785398}$     &       $-17.20$          &      $-131.1$       &          $-1793$       \\ 
${\bf -0.785398}$     &       ${\bf -16.9356}$          &      $-175.9$       &          \multicolumn{1}{c}{--}       \\ 
${\bf -0.785398}$     &       ${\bf -16.9356}$          &      $-174.4$       &          $-107$       \\ 
\hline
\end{tabular}
\caption{
\label{Table:fit_cons}
Values of the PM coefficients $a_n$ obtained by fitting the numerical data for $\delta\chi^{\rm cons}$ to the PM model (\ref{model_cons}) for the orbits in Table~\ref{Tablev50}, i.e.\ ones with $v=0.5$ and $60\leq b \leq 125$. In each row, non-empty entries represent terms fitted for, except entries in bold, which are fixed at their known PM values, given by Eqs.~(\ref{eq:chi_dis_3PM}) and (\ref{eq:chi_dis_4PM}) for $a_3$ and $a_4$ respectively.
}
\end{center}
\end{table}

Next consider the fitted values for the unknown terms $a_4$ and $a_5$. Table \ref{Table:fit_cons} shows the different values obtained with different choices of $\bar n$ and of whether $a_2$ and $a_3$ are fitted for or fixed at their known values (bold entries in the table denote fixed analytical values). The 4PM coefficient $a_4$ appears to settle at around $\sim\! -175$ (for $v=0.5$), but the results for 5PM are manifestly unstable, suggesting our data are insufficient for estimating the 5PM term. (Recall that, for simplicity, we are not accounting here for $\log(b)$-running terms, which, at any rate, produce a very small variation over the range of $b$ values in our sample.)

Focusing now on the 4PM term, and recalling Eq.~(\ref{eqn:4PMCons}), we write it in the form 
\begin{equation}
\delta\chi_4^{\rm cons} = \delta\chi_4^{\rm known}+\Delta_4(v)/b^4\,,
\end{equation}
where,
\begin{align}
\label{Delta_4_form}
    \Delta_4(v) :=& \: \frac{3}{8}\frac{\pi G^3 m_2^4}{(\sigma^2-1)} 
                    \left(c_1 (\sigma^2-5) + c_2(\bar \mu) (\sigma^2-1) \right) \nonumber
\\
                =& \: \frac{3}{8}\pi G^3 m_2^4 \left[c_2(\bar\mu)+c_1(5-4/v^2)\right]   
\end{align}
is the Wilson-coefficients contribution to $\delta\chi_4^{\rm cons}$, divided by $b^4$. The parameters $c_1$ and $c_2(\bar \mu)$ are a priori unknown. As mentioned previously we expect $c_1 = 0$, but leave it for comparison to the SF results. As commented above, in quantum field theory, the process of renormalization introduces a running scale $\bar\mu$ for the  $c_2$ coefficient, which we choose $\bar\mu= (2 G m_2)^{-1}$. For brevity, we suppress the scale dependence below.
Thus we have 
\begin{equation}\label{eqn:EpilsonDefn}
\Delta_4(v;c_1,c_2)= b^4\left(\delta\chi^{\rm cons} - \delta\chi^{\rm cons}_2 - \delta\chi^{\rm cons}_3 - \delta\chi^{\rm cons}_4 \big|_{c_1=c_2=0}\right)  +\calO(1/b).
\end{equation}
We can construct the right-hand side of (\ref{eqn:EpilsonDefn}) by subtracting the known PM terms from the numerical data, and then attempt to fit the residual with the expected form shown in Eq.~(\ref{Delta_4_form}), in  order to find $c_1$ and $c_2$. 

\begin{figure}[ht!]
\centering
\includegraphics[width=.8\linewidth]{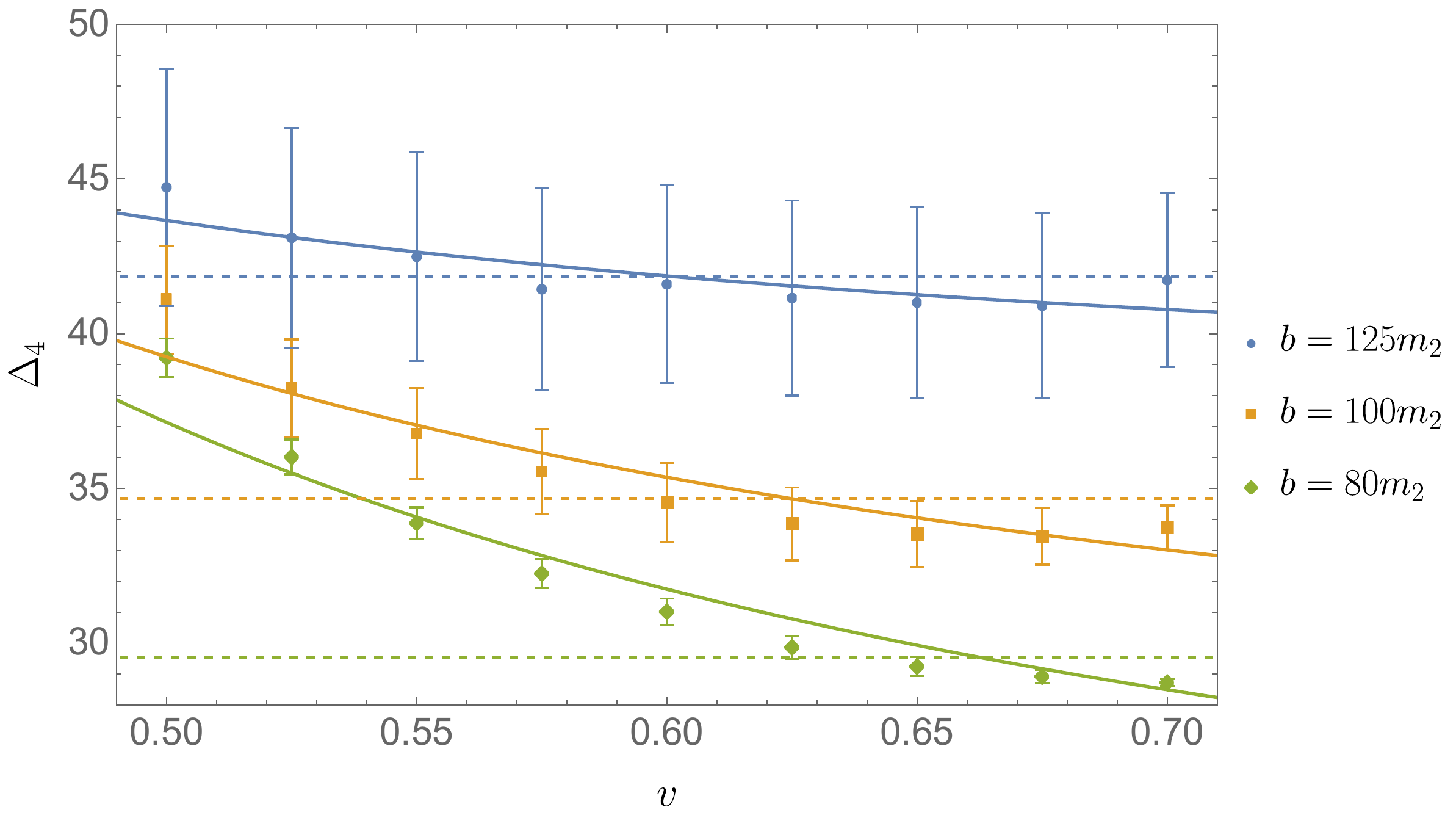}
\caption{
The 4PM residue $\Delta_4$ as a function of velocity for 3 constant values of impact parameters $b[=(80,100,125)m_2]$. The data points are the values (and error bars) of $\Delta_4$ as determined from the numerics via Eq.\ (\ref{eqn:EpilsonDefn}). The solid lines are least-squares fits to the model described in the second line of Eq.\ (\ref{Delta_4_form}). The dashed lines are similar fits where we have fixed $c_1=0$.}
\label{Cons4PMFit}
\end{figure}

For that purpose, we have prepared three sets of data, each with fixed $b[=(80,100,125)M]$ and varying $v$. Figure \ref{Cons4PMFit} shows our fitted functions $\Delta_4(v)$, superposed on the numerical data, for each of the three fixed values of $b$. We note immediately the relatively large numerical error bars on $\Delta_4$ in the plot (which, recall, constitutes the very small residual left after subtracting all known PM terms). The noisiness of the data is clearly visible, especially at $b=125M$, and we expect it to restrict the accuracy of our fit.  Also of notice is the larger-than-expected variation in $\Delta_4(v)$ as a function of $b$. We expect $\Delta_4(v)$ to converge as $\sim 1/b$ for $b\to\infty$, but the results suggest we are not quite yet in a convergent regime---possibly due to large contributions from omitted 5PM terms. This too warns us that our fitted values for $c_1$ and $c_2$ might not be as reliable as we might have hoped.

Our best-fit results for $c_1$ and $c_2$ are presented in the first 3 lines of Table \ref{c1&c2}. In the table, error bars are (least-squares) model fitting errors, and do not directly take into account the data error bars displayed in Fig.~\ref{Cons4PMFit}. As expected, the fitted values of $c_1$ and $c_2$ carry sizable error bars, especially at larger $b$ where the quality of $\Delta_4(v)$ data is poorer. Moreover, the values obtained with each of the three fits do not appear to be consistent with each other, not even within their large error bars. We must conclude that we have insufficient data to extract the two unknown coefficients $c_1$ and $c_2$ individually. To enable this, we need more accurate data sampled at larger values of $b$.

\begin{table}[h]
\begin{center}
\begin{tabular}{c|c|c}
\hline
$b/m_2$ & $c_1$ & $c_2$  \\ \hline
$80$ & $0.94$ & $-21.2$ \\
$100$ & $0.68$ & $-25.9$ \\
$125$ & $0.31$ & $-33.6$ \\
$80$ & ${\bf 0}$ & $-25.1$ \\
$100$ & ${\bf 0}$ & $-29.4$ \\
$125$ & ${\bf 0}$ & $-35.5$ \\
\hline
\end{tabular}
\caption{
\label{c1&c2}
Values of the coefficients for $c_1$ and $c_2$ for the fits shown in Fig.\ \ref{Cons4PMFit}. Entries in bold are fixed to $c_1=0$, hypothesized based on some theoretical evidence.  Statistical fitting errors are significantly smaller than the data errors shown in Fig.\ \ref{Cons4PMFit}, so we do not display them here.  However, it is worth noting the large fitting error in the value of $c_1$ in the third row, $c_1 = 0.31\pm0.38$, making this coefficient statistically consistent with zero.  
}
\end{center}
\end{table}

As discussed in Sec.~\ref{subsec:EFT_and_tidal}, there are compelling theoretical reasons to expect $c_1=0$. Taking this value, $\Delta_4$ becomes a ($v$-independent) constant number.  Unfortunately, it remains unclear whether our data supports this expectation. In Fig.\ \ref{Cons4PMFit}, the data for $b=80M$ and $b=100M$ shows a variation of $\Delta_4$ with $v$, although a leveling-off is evident at large $v$ (corresponding to weaker-field orbits). We suspect the smaller-$v$ portion of these two datasets contains a large 5PM contribution, which diminishes as $v$ is increased (hence also increasing $r_{\rm min}$). The data for our weakest-field set, with $b=125M$, shows a greater consistency with $c_1 = 0$, but also suffers from larger numerical error.  The last 3 rows of Table \ref{c1&c2} display the outcome of fitting for $c_2$ while holding $c_1$ fixed at zero, and the corresponding (constant) values of $\Delta_4$ are also shown in Fig.~\ref{Cons4PMFit}. 

\begin{figure}[h!]
\centering
\includegraphics[width=.8\linewidth]{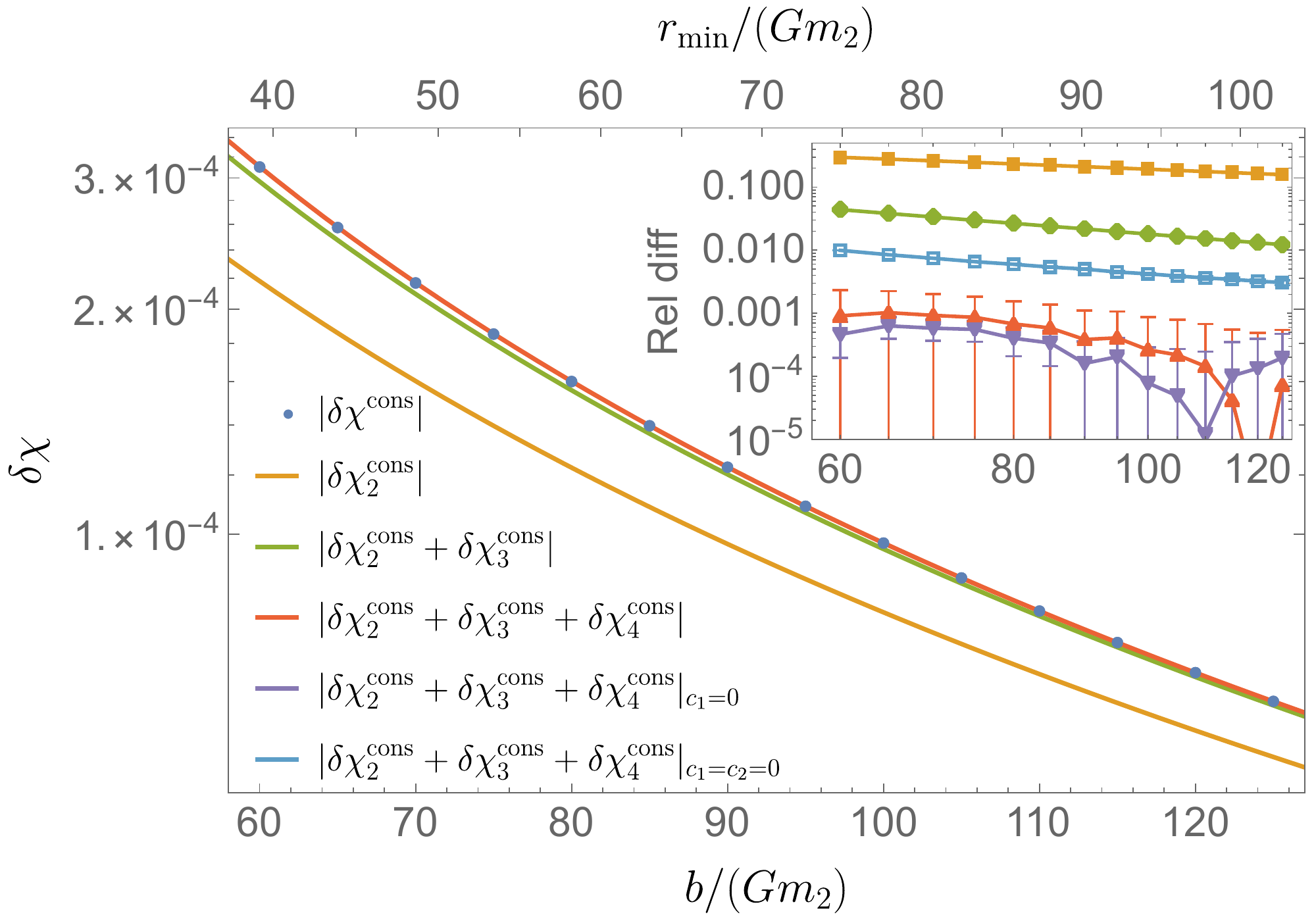}
\caption{
The conservative self-force correction to the scattering angle for the values shown in Table \ref{Tablev50}, with $v=0.5$. The numerical data points are marked in blue. The various analytical PM approximations are shown in solid lines: In the main plot we show the 2PM result (orange), 3PM result (green) and 4PM result (red), whereas in the latter we have used the best-fit values of the a-priori unknown 4PM coefficients $c_1$ and $c_2$ from the 3rd row of Table \ref{c1&c2}.
The inset shows the relative difference between the numerical data and the various PM approximations. For the 4PM approximation we show three alternative models, corresponding to (1) the best-fit values of $c_1$ and $c_2$ (red line, as in the main plot); (2) forcing $c_1=0$ and using the best-fit value for $c_2$ from the 6th row of Table \ref{c1&c2} (purple); and (3) setting $c_1=0=c_2$ (light blue, as a reference). 
}
\label{v50ConsPMComp}
\end{figure}

Finally, Fig.~\ref{v50ConsPMComp} summarizes our findings for the conservative sector. It shows the numerical data points (as a function of $b$ at fixed $v=0.5$) together with the various PM approximations up to 4PM. In the 4PM case we present three variations: (1) The 4PM result with $c_1$ and $c_2$ both set to zero; (2) the 4PM result with best-fit values for $c_1$ and $c_2$ from the third line of Table \ref{c1&c2}; and (3) the 4PM result with $c_1=0$ and the best-fit value for $c_2$ from the sixth line of Table \ref{c1&c2}.  It is striking to observe that, with the best-fit value for the unknown parameters (and especially when forcing $c_1=0$), the 4PM model agrees with the full SF data to within a few parts in $10^4$ over the entire range of orbital parameters in our sample.

\subsection{Dissipative sector: fitting for $\delta\chi^{\rm diss}_{>4}$}

We have obtained $\delta\chi^{\rm diss}$ through 4PM order, but no terms are currently known at 5PM or beyond. Here we will attempt to extract information about higher-order terms by fitting our numerical data to a power series of the form 
\begin{equation}
\label{model_dis}
\delta\chi^{\rm diss}(v,b) = \sum_{n=3}^{\bar n} \frac{\alpha_n(v)}{b^n} + O\left(b^{-\bar n-1}\right).
\end{equation}
The coefficients $\alpha_3\equiv\delta\chi^{\rm diss}_3$ and $\alpha_4\equiv\delta\chi^{\rm diss}_4$ are the known ones: they are given in Eqs.\ (\ref{eq:chi_dis_3PM}) and (\ref{eq:chi_dis_4PM}), respectively. The coefficients $\alpha_{n>4}$ are not known. Once again, we will fit with a range of truncation orders $\bar n$ to provide some control over the quality of the fit, and we will first use the data to fit for the {\it known} PM terms as a test. 

The results are shown in Table \ref{5PMTable}, which is arranged in the style of Table \ref{Table:fit_cons}.
In the first 4 rows, no analytical knowledge is assumed, and we try to fit the data (at fixed $v=0.5$) to the model (\ref{model_dis}) with $\bar n=3$ (first row), $\bar n=4$ (second row), $\bar n=5$ (third row) and $\bar n=6$ (fourth row). We see that the fitted value of $\alpha_3$ settles to a value within a fraction of a percent of its analytically predicted value,  $\alpha_3\sim\!9.6225$. The fitted value of the NLO coefficient, $\alpha_4$, is less stable, but varies within a few percent of the analytical value, $\alpha_4\sim 143$. This remains the case also when the fit is performed with the analytical value of $\alpha_3$ fixed (fifth to seventh rows). These experiments provide further reassurance about the validity of our analytical results at 3PM and 4PM.

\begin{table}[h]
\begin{center}
\begin{tabular}{l|l|r|r}
\hline
\multicolumn{1}{c|}{$\alpha_3$}            & \multicolumn{1}{c|}{$\alpha_4$}           & \multicolumn{1}{c|}{$\alpha_5$}       & \multicolumn{1}{c}{$\alpha_6$}           \\ \hline
$11.19$     &       \multicolumn{1}{c|}{--}          &      \multicolumn{1}{c|}{--}       &          \multicolumn{1}{c}{--}       \\ 
$9.44$   & $188$       &       \multicolumn{1}{c|}{--}      &        \multicolumn{1}{c}{--}         \\ 
$9.64$ & $142$   & $1900$ &         \multicolumn{1}{c}{--}        \\ 
$9.61$  & $154$   & $920$  & $26615$   \\ 
${\bf 9.6225}$         & $169$   &       \multicolumn{1}{c|}{--}      &        \multicolumn{1}{c}{--}         \\ 
${\bf 9.6225}$         & $147$ & $1720$  &       \multicolumn{1}{c}{--}          \\ 
${\bf 9.6225}$         & $149$  & $1321$ & $15859$ \\ 
${\bf 9.6225}$         & ${\bf 143.344}$       & $1965$ &     \multicolumn{1}{c}{--}            \\ 
${\bf 9.6225}$         & ${\bf 143.344}$       & $2248$ & $-20216$ \\ \hline
\end{tabular}
\caption{
\label{5PMTable}
Values of the PM coefficients $\alpha_n$ obtained by fitting the numerical data for $\delta\chi^{\rm diss}$ to the PM model (\ref{model_dis}) for the orbits shown in Fig.~\ref{orbits}, i.e.\ ones with $v=0.5$ and $60\leq b \leq 125$. In each row, non-empty entries represent terms fitted for, except entries in bold, which are fixed at their known PM values, given by Eqs.~(\ref{eq:chi_dis_3PM}) and (\ref{eq:chi_dis_4PM}) for $\alpha_3$ and $\alpha_4$ respectively.
}
\end{center}
\end{table}

We next consider the fitted values for the unknown terms $\alpha_5$ and $\alpha_6$. Table \ref{5PMTable} shows the different values obtain with different choices of $\bar n$ and of whether $\alpha_3$ and $\alpha_4$ are fitted for or fixed at their known values (recall bold entries denote fixed analytical values). The 5PM coefficient $a_5$ appears to vary around $\sim\!2000$ (for $v=0.5$), with an uncertainty of several hundred. The results for 6PM appear to be completely unstable, implying our data is insufficient for estimating the 6PM term. 

\section{Conclusions and outlook }
\label{sec:Outlook}

Recent years have seen a major effort to produce new results in both the SF and the PM approaches to the two-body problem.   In particular, results for the second-order SF are now available for the bound systems~\cite{Wardell:2021fyy}, and the first results for scattering processes are now becoming available~\cite{Long:2021ufh, Barack:2022pde}.  There have also been analogous strides in the PM expansion, with results available through $\mathcal O (G^4)$~\cite{Bern:2021dqo, Bern:2021yeh,  Manohar:2022dea, Dlapa:2021npj, Dlapa:2021vgp, Dlapa:2022lmu,Dlapa:2023hsl}.

In this paper, we carried out an initial precision comparison between SF and PM scattering angles.
We did so in the context of a scalar-field model~\cite{Gralla:2021qaf, Barack:2022pde} as a proof of principle for the analogous comparison in the purely gravitational problem. Scattering processes facilitate comparisons between different approaches because they involve physical observables defined at infinity. Such cross-checks between different frameworks in overlap regions will be crucial in the future, not only to help demonstrate the reliability of each approach but also to synthesize improved approximation schemes valid beyond the reach of each approach alone. This is especially important for extreme mass-ratio systems which are difficult to analyze with numerical relativity methods. 
The semi-analytical nature of the SF approach allows for the separation of contributions from scalar and gravitational back reactions and of conservative and dissipative effects, providing a means for detailed comparisons.

We considered a two-body scattering process and compared the scattering angle through $\calO(G^3 q_s)$ and leading SF order, leaving for future studies comparisons of other interesting quantities, such as the energy and angular momentum fluxes.
The SF results used in our analysis were obtained using the methods from Ref.~\cite{Long:2021ufh} together with improvements in error mitigation and analysis.  
The corresponding PM results were obtained via the methods of Refs.~\cite{Cheung:2018wkq, Bern:2019nnu, Bern:2019crd, Bern:2021dqo, Bern:2021yeh, Manohar:2022dea}.  
\enlargethispage{\baselineskip}
We directly compared the perturbative PM and SF results finding excellent agreement of a few parts in $10^{4}$ in the regime where we expect that both SF and PM calculations are valid.

A feature encountered in the PM expansion of the scalar model is the appearance of an operator  with tidal effects resulting in an ultraviolet divergence if the operator is ignored.  This phenomenon occurs at three loops in the scalar-field model, but it is delayed until five loops in the purely gravitational problem, due to additional derivatives present in gravitational couplings compared to scalar interactions. 
The extra derivatives raise the dimension of corresponding tidal operators, pushing their appearance to higher orders.  
The main consequence of this operator, which foreshadows the analogous one in the purely gravitational problem, is that there are two (counterterm) coefficients that must be determined through a matching calculation.  One of these coefficients is expected to be zero~\cite{Fang:2005qq, Damour:2009vw, Binnington:2009bb, Kol:2011vg, Landry:2015zfa, LeTiec:2020bos, Chia:2020yla, Hui:2020xxx, Charalambous:2021mea, Hui:2021vcv, Ivanov:2022qqt, Ivanov:2022hlo}, compatible also with our comparison to SF, but some care is needed before concluding this since coefficients can in principle be shifted from other evaluations by scheme choices and field redefinitions.  Here, we did not carry out the required matching calculation; instead, we allowed these parameters to float freely when aligning the 1SF and 4PM results for the scattering angle. 
It would also be useful to carry out the matching calculation to determine the coefficients $c_1$ and $c_2$ as preparation for future 6PM calculations in the purely gravitational case, for which contributions of tidal operators to the scattering angle are delayed until that order. 

Various subtleties arise at high orders in both the SF and PM approach.  While scattering processes simplify the definition of gauge invariant asymptotic observables (e.g. scattering angles or impulses) similar comparisons should be carried out for the bound state problems.  However, at the 4PM order and beyond, analytic continuations from unbounded to bound systems in General Relativity are no longer straightforward~\cite{Damour:2014jta, Bini:2017wfr, Bini:2020nsb, Dlapa:2021vgp}, due to non-local-in-time effects; this issue needs to be resolved for precision  bound-state comparisons to be carried out between PM and SF calculations.
Another subtlety is that at higher orders in SF the separation of conservative and dissipative effects becomes definition-dependent, so direct comparisons across formalisms could help shed light on this. Another important direction is to incorporate any PM scattering results into an EOB framework which greatly enhances its region of validity, as carried out in General Relativity at 4PM in  Refs.~\cite{Khalil:2022ylj, Damour:2022ybd}.  

A key goal is to carry out similar comparisons for two black holes interacting purely gravitationally, instead of the scalar model used here, using EOB-improved PM calculations.  We look forward to future SF and PM calculations that will allow detailed comparisons and the construction of EOB and other models valid for extreme mass ratios and to higher precision than currently possible.

\subsection*{Acknowledgements}
We thank Misha Ivanov, Rafael Porto, Zihan Zhou for useful discussion. We are especially grateful to Thibault Damour for detailed discussions on the mass polynomiality of the impulse. We would also like to thank Benjamin Leather for providing the code used to generate Figure~\ref{TwoBodyDiagram}.
 This research was supported in part by the U.S. Department of Energy (DOE) under award numbers DE-SC0009937, DE-SC00019066, DE-SC0011632 and DE-SC0009919 and in part by the U.S. National Science Foundation under Grant No. NSF PHY-1748958.  In relation to the latter support, we thank the Kavli Institute for Theoretical Physics for their hospitality, where this paper was initiated. 
 M.Z.'s work is supported in part by the U.K. \ Royal Society through Grant URF\textbackslash R1\textbackslash 20109. For the purpose of open access, the author has applied a Creative Commons Attribution (CC BY) license to any Author Accepted Manuscript version arising from this submission.
J.P.-M. would like to thank the Institut des Hautes \'{E}tudes Scientifiques and the Korea Institute for Advanced Study, for their hospitality while this work was being completed. M.P.S. is supported by the Alfred P. Sloan Foundation. We are also grateful to the Mani L. Bhaumik Institute for Theoretical Physics and the Walter Burke Institute for Theoretical Physics for support. We acknowledge the use of the IRIDIS High-Performance Computing Facility, and associated support services at the University of Southampton, in the completion of this work. This work makes use of the Black Hole Perturbation Toolkit \cite{BHPToolkit}. 

\appendix
\section{Lorentz Frames}
\label{sec:Frames}

In this appendix, we show that the scalar-field SF calculation yields the same result in several common choices of inertial frame. For instance, one can set up the calculation in the initial rest frame of the heavy black hole (scalar $\phi_2$), which is common in PM calculations. But the SF calculation is often done in the final rest frame of the black hole. Alternatively, one can also choose the center-of-mass frame of the initial or final state. As we will see, these choices all lead to the same scalar-field SF results at $\mathcal{O}(q_s)$. Note, however, they would lead to different {\em gravitational} SF corrections, which are of order $\mathcal{O}(q_m)=\mathcal{O}(m_1/m_2)$.

Let us begin by recalling the structure of the scattering angle under PM expansion,
\begin{align}
	\chi &\sim \sum_{n=1} \, \left(\frac{G m_2}{b}\right)^n (c_{n,0}
    + d_{n,1}\, q_m )
    +\sum_{n=1} \, \left(\frac{G m_2}{b}\right)^{n-1} q_s c_{n,1}
    +\dotsc,
	\label{eq:angle_SFcounting}
\end{align}
where $c_{n,0}$ and $c_{n,1}$ are coefficients at $n$PM order describing the geodesic and first-order SF effects. For completeness, we also include the first-order gravitational SF correction, whose coefficients are $d_{n,1}$. Recall the definitions of $q_m$ and $q_s$ in Eqs.~\eqref{eq:qm_def} and \eqref{eq:qs_def}. The ellipsis stands for higher orders in $q_s$ and/or $q_m$.

Since we consider SF corrections, we expect the heavy black hole remains non-relativistic in all the inertial frames mentioned earlier, i.e. the rest/center-of-mass frame in the initial or final state. Therefore we can estimate the effect by a non-relativistic boost. First, consider boosting from the center-of-mass frames to the rest frame of the heavy black hole, either in the initial or final states. The boost parameter is the same order as the velocity of the heavy black hole, $|\bm p_2|/m_2$. 
\begin{align}
    v \sim \frac{|\bm p_2|}{m_2} &\sim \frac{|\bm p_1|}{m_2} \sim \frac{m_1}{m_2} = \mathcal{O}(q_m),
    \label{eq:CM_boost}
\end{align}
where we use $|\bm p_2|\sim |\bm p_1|$ in the center-of-mass frame and $|\bm p_1|$ is approximately $m_1$ times a numerical factor that depends on the boost factor $\sigma$. Applying the boost to the scattering angle yields the correction $\sim v\times \chi$. Combining Eqs.~\eqref{eq:angle_SFcounting} and \eqref{eq:CM_boost}, we see that the boost only changes the coefficients $d_{n,1}$, and does not affect the geodesic or scalar-field SF results, $c_{n,0}$ and $c_{n,1}$ respectively. One can consider the SF correction to $|\bm p_1|$, but this only results in higher orders in $q_s$ or $q$. We conclude that the first-order scalar-field SF correction to the scattering angle is the same in the rest frame on $m_2$ as it is in the center-of-mass frame.

We consider next the difference between the initial and the final rest frames of the black hole. The difference is due to the recoil on the black hole. We can estimate the velocity by
\begin{align}
    v \sim \frac{|\Delta \bm p_2|}{m_2}.
\end{align}
In the conservative case, we can use $\Delta \bm p_2 = -\Delta \bm p_1$ from momentum conservation. Combining with $|\Delta \bm p_1|\sim |p_1| \chi$ yields
\begin{align}
    v \sim \frac{|\Delta \bm p_2|}{m_2} \sim \frac{|\Delta \bm p_1|}{m_2} \sim \frac{m_1}{m_2} \chi = \mathcal{O}(q_m).
\end{align}
Again, applying this boost only affects angles as $v\times \chi$, so the scalar-field SF effect remains the same. 

In the presence of dissipation, the impulse $\Delta \bm p_2$ receives contributions from the radiated momentum and, as far as the SF counting is concerned, it obeys $|\Delta \bm p_2| \sim |\bm P_{\rm rad}| \sim E_{\rm rad}$. We observe that
\begin{align}
    E_{\rm rad} \sim \frac{m_1 m_2}{b} \left(\frac{G m_2}{b}\right)^2 q_s e_{3,1} + 
    m_1 \left(\frac{G m_2}{b}\right)^3 q_m \tilde{e}_{3,1} +\dots,
\end{align}
which can come from scalar-field and gravitational radiation, respectively. 
We can see that $E_{\rm rad}/m_1$ is suppressed by $q_s$ or $q_m$, since there is no energy loss in the geodesic limit.
The corresponding boost velocity is
\begin{align}
    v \sim \frac{|\Delta \bm p_2|}{m_2} \sim \frac{E_{\rm rad}}{m_2} \sim \mathcal{O}(q_m).
\end{align}
We, therefore, conclude that, when passing from the rest frame to the 
center-of-mass frame, the scalar-field SF correction to the scattering angle is unchanged by passing from the initial rest frame to the final one.


\newpage

\bibliographystyle{JHEP}
\bibliography{refs.bib}

\end{document}